\documentclass[lettersize,journal]{IEEEtran}

\usepackage[english]{babel}

\usepackage{amsmath,amsfonts,amssymb}
\usepackage{algorithmic}
\usepackage{algorithm}
\usepackage{array}
\usepackage[caption=false,font=normalsize,labelfont=sf,textfont=sf]{subfig}
\usepackage{textcomp}
\usepackage{stfloats}
\usepackage{verbatim}
\usepackage{cite}
\usepackage{mathbbol}

\usepackage{tcolorbox}
\tcbuselibrary{listings,theorems}

\DeclareSymbolFontAlphabet{\mathbbm}{bbold}
\DeclareSymbolFontAlphabet{\mathbb}{AMSb}%

\usepackage{mathrsfs}  
\usepackage{graphicx}
\usepackage[colorlinks=false]{hyperref}
\usepackage{bm}
\usepackage{tikz}
\usetikzlibrary{shapes.geometric, arrows,fit,shapes.arrows,calc,positioning}
\tikzstyle{arrow} = [thick,->,>=stealth]
\tikzstyle{box} = [rectangle, rounded corners, minimum width=5cm, text width=4cm, minimum height=3cm,align=left, draw=black, fill=red!30]
\DeclareMathAlphabet{\mathscrbf}{OMS}{mdugm}{b}{n}

\newcommand{\N}{\mathbb{N}}
\newcommand{\Z}{\mathbb{Z}}
\newcommand{\R}{\mathbb{R}}
\newcommand{\C}{\mathbb{C}}

\newcommand{\eUnit}{\mathrm{e}}
\newcommand{\iUnit}{\mathrm{i}}
\newcommand{\mo}{m_0}
\newcommand{\MS}{M_\mathrm{S}}
\newcommand{\kB}{k_\mathrm{B}}
\newcommand{\TP}{T_\mathrm{P}}
\newcommand{\Vc}{V_\mathrm{c}}
\newcommand{\vek}[1]{\bm{#1}}
\newcommand{\Kanis}{K^\text{anis}}
\newcommand{\Kanismax}{K^\text{anis}_\text{max}}
\newcommand{\isoParaSet}{\mathbb{O}_{\mathscr{L}}}

\newcommand{\dndt}[1][t]{\frac{\mathrm{d}}{\mathrm{d}#1}}

\newcommand{\pnpt}[2][t]{\frac{\partial #2}{\partial #1}}
\newcommand{\pnptn}[2][t]{\frac{\partial}{\partial #1}\!\left[#2\right]}
\newcommand{\dt}[1][t]{\,\mathrm{d}#1}

\newcommand{\mbar}[1]{\bar{\bm{m}}\!\left(#1\right)}
\newcommand{\tran}{\intercal}
\newcommand{\ak}{\alpha_\mathrm{K}}
\newcommand{\HD}{\bm{H}^\mathrm{D}}
\newcommand{\HS}{\bm{H}^\mathrm{S}}
\newcommand{\TD}{T_\mathrm{D}}
\newcommand{\NB}{N_\mathrm{B}}
\newcommand{\fB}{f_\mathrm{B}}
\newcommand{\rect}{\operatorname{rect}}
\newcommand{\sgn}{\operatorname{sgn}}

\usepackage{color}
\definecolor{ibilight}{RGB}{193,216,237}
\definecolor{ibidark}{RGB}{0,73,146}	
\definecolor{uke2}{RGB}{170,156,143} 	
\definecolor{uke3}{RGB}{87,87,86}		
\definecolor{ukesec1}{RGB}{255,223,0}	
\definecolor{ukesec2}{RGB}{239,123,5}	
\definecolor{ukesec3}{RGB}{104,195,205}	
\definecolor{ukesec4}{RGB}{138,189,36}	
\definecolor{ukesec5}{RGB}{178,34,41}	
\definecolor{tuhh}{RGB}{45,198,214}     
\definecolor{ibidarkBG}{RGB}{227,229,242}   
\definecolor{uke2BG}{RGB}{233,228,225} 	    
\definecolor{uke3BG}{RGB}{230,231,232}	    
\definecolor{ukesec1BG}{RGB}{255,243,190}   
\definecolor{ukesec2BG}{RGB}{254,232,212}   
\definecolor{ukesec3BG}{RGB}{222,241,241}   
\definecolor{ukesec4BG}{RGB}{233,243,222}   
\definecolor{ukesec5BG}{RGB}{244,230,225}   

\newtcbtheorem[auto counter, number format=\Roman]{tlemma}{Lemma}{colback=uke2BG,colframe=uke2}{th}

\title{Equilibrium Model with Anisotropy for Model-Based Reconstruction in Magnetic Particle Imaging }

\author{Marco Maass, Tobias Kluth, Christine Droigk, Hannes Albers, Konrad Scheffler, Alfred Mertins, Tobias Knopp
\thanks{
H. Albers and T. Kluth acknowledge funding by the German Research Foundation (DFG, Deutsche Forschungsgemeinschaft) - project 426078691.}
\thanks{
M. Maass is with the German Research Center for Artificial Intelligence (DFKI), 23562 Lübeck, Germany (e-mail: marco.maass@dfki.de).
M. Maass, C. Droigk, and A. Mertins  are with the Institute for Signal Processing, University of Lübeck, 23562 Lübeck, Germany (e-mail: c.droigk@uni-luebeck.de,  alfred.mertins@uni-luebeck.de).
H. Albers and T. Kluth are with the Center for Industrial Mathematics, University of Bremen, 28359 Bremen, Germany (e-mail: halbers@uni-bremen.de, tkluth@math.uni-bremen.de).
T. Knopp and K. Scheffler are with the Section for Biomedical Imaging, University Medical Center Hamburg-Eppendorf, 20246 Hamburg, Germany and the Institute for Biomedical Imaging, Hamburg University of Technology, 21073 Hamburg, Germany (e-mail: t.knopp@uke.de). T. Knopp is also with the Fraunhofer Research Institution for Individualized and Cell-Based Medical Engineering IMTE, 23562 Lübeck, Germany.}
}

\begin{document}

\maketitle

\begin{abstract}
Magnetic particle imaging is a tracer-based tomographic imaging technique that allows the concentration of magnetic nanoparticles to be determined with high spatio-temporal resolution. To reconstruct an image of the tracer concentration, the magnetization dynamics of the particles must be accurately modeled.  A popular ensemble model is based on solving the Fokker-Plank equation, taking into account either Brownian or Néel dynamics. The disadvantage of this model is that it is computationally expensive due to an underlying stiff differential equation.
A simplified model is the equilibrium model, which can be evaluated directly but in most relevant cases it suffers from a non-negligible modeling error. In the present work, we investigate an extended version of the equilibrium model that can account for particle anisotropy. We show that this model can be expressed as a series of Bessel functions, which can be truncated based on a predefined accuracy, leading to very short computation times, which are about three orders of magnitude lower than equivalent Fokker-Planck computation times. We investigate the accuracy of the model for 2D Lissajous magnetic particle imaging sequences and show that the difference between the Fokker-Planck and the equilibrium model with anisotropy is sufficiently small so that the latter model can be used for image reconstruction on experimental data with only marginal loss of image quality, even compared to a system matrix-based reconstruction. 

\end{abstract}


\section{Introduction}

\IEEEPARstart{T}{he} tomographic imaging method magnetic particle imaging (MPI) uses superparamagnetic iron oxide nanoparticles (SPIOs) and determines their concentration in space and time \cite{Gleich2005}. MPI has a high spatio-temporal resolution and can image even smallest amounts of tracer concentrations \cite{graeser2017towards}. 
Although MPI is still in the preclinical phase, there are a growing number of potential clinical applications, for example in vascular imaging \cite{herz2017magnetic,kaul2018magnetic,vogel2020superspeed, tong2023atherosclerosis}, perfusion imaging \cite{ludewig2017magnetic,molwitz2019first,szwargulski2020monitoring}, cancer imaging \cite{zhu2019quantitative,huang2023glioblastoma}, and interventional imaging \cite{salamon2016magnetic,rahmer2017interactive,herz2019magnetic}.

During an imaging experiment, the MPI scanner applies a highly dynamic and spatially varying magnetic field sequence to encode information about the  concentration in the resulting particle magnetization. The magnetization change is recorded using one or multiple receive coils. The goal is then to reconstruct the particle concentration from the received signal, which is a classical linear inverse problem that requires proper regularization to cope with the ill-posedness of the imaging equation \cite{maerz2015modelbased,Kluth2018b, erb2018mathematical}.

One important precondition for inverse problems in general and also for image reconstruction in MPI is that the forward operator mapping from particle concentration to recorded voltage signals is precisely known. 
Modeling the entire signal acquisition chain includes several scanner components as well as the tracer material. While scanner components like the exact applied field can be determined with moderate calibration effort \cite{thieben2022efficient}, it remains a key challenge to provide an adequate particle magnetization model. 
The magnetization of SPIO distributions is physically difficult to model accurately \cite{Weizenecker2012MicroMagneticSimulation,Kluth2018a, Weizenecker2018}, and it can be computationally challenging to accurately predict the particle parameters \cite{albers2022simulating}. 
This holds particularly true when one takes into account Lissajous-type excitation allowing for short measurement times. To circumvent these issues, the state of the art procedure for obtaining a discrete version of the forward operator (i.e. the system matrix) exploits a small calibration sample that is moved to a predefined set of positions in space sequentially, measuring the system matrix explicitly. This procedure is time-consuming requiring hours to days even for small grid sizes and it, in particular, does not generalize with respect to scanner as well as particle parameters  \cite{Knopp2017}. 

Model-based considerations to predict system performance already go back to the very beginning of MPI research \cite{rahmer2009signal} (see also the survey articles \cite{Knopp2017,Kluth2018a}).
Model-based reconstruction approaches were then first investigated in \cite{knopp2009model,Knopp2010d} for one-dimensional as well as multi-dimensional excitation for a simplified particle model assuming equilibrium (equilibrium model, EQ). One-dimensional excitation strategies, which typically come up with a significant increase of measurement times, are essential for the $x$-space reconstruction approach \cite{Goodwill2010} but may show a less complex magnetization behavior, which has been tackled using the Debye model \cite{croft2012relaxation}. The results for multi-dimensional excitation have been significantly improved by introducing a Fokker-Planck (FP) model, which involves solving a differential equation and can both describe Néel and Brown dynamics (e.g., see \cite{Kluth2018a,albers2022simulating}). It has been shown that the model can describe both fluid particles \cite{Kluth2019} and immobilized particles either with \cite{albers2022modeling} or without \cite{Albers2023IWMPI} alignment. 

Despite this recent progress, finding model parameters (i.e. particle core diameter and particle anisotropy distributions) remains a challenge because the FP model is computationally expensive and thus limits the parameter space that can be explored. One first attempt in the direction of polydisperse parameter identification has been made in \cite{albers2022simulating} and more recently, the models were combined with a measured transfer function of the receive chain \cite{Albers2023IWMPI}, which allowed  to determine more physically meaningful particle model parameters.
To speed up computation, quite recently two  promising alternatives to the FP model were proposed. The first is the approximation of the FP model by a neural operator \cite{Knopp2023IWMPI}. The other is the extension of the equilibrium model with anisotropy (EQANIS) \cite{albers2022immobilized,maass2022analytical}, which makes the model much more expressive than the EQ model. 

The contributions of this paper are manifold. We provide the complete mathematical description and derivation of the EQANIS model, which was only sketched in the previous conference papers. This includes the derivation of an efficient computational strategy for the EQANIS model. In a one-dimensional excitation MPI simulation study for different SPIO parameters, the limits of the approximation performance of the EQANIS model to the Néel-FP model are systematically investigated, so that other researchers can check whether the EQANIS model is suitable for their purposes. In addition, the number of summands required to obtain a small numerical truncation error as a function of the SPIO parameters in the EQANIS model is investigated. This allowed us to significantly reduce the number of summands compared to previous work, such that the evaluation time is almost constant and independent of the particle parameters, which is not the case in the Néel-FP model. As a result, we can show that the model-based MPI system function can be computed two to three orders of magnitude faster with the EQANIS model than with the Néel-FP model.
In addition, we have extended the underlying model B3 of \cite{Kluth2019} by a non-linear parameter, which improves the model accuracy for both the Néel-FP model and the EQANIS model. 
Since the EQANIS model has only been considered in simulations so far, another main purpose of this work is to evaluate the model on experimental data and to investigate its suitability for model-based reconstruction, supporting the theoretical considerations with experimental results.
Following the principles of open science, we make the source code of the EQANIS model and all experimental datasets used in this work openly available so that they can be freely used for future research.

\section{Physical Models}

Modeling the entire signal acquisition chain in MPI requires the consideration of various components. 
From a very general point of view, one can distinguish two main model building blocks: the scanner model and the particle model (more precisely, particle magnetization model). The scanner model mainly includes the applied magnetic field, the receive coil sensitivity, and the analog filter prior to digitization. 
The particle model includes the magnetization of large nanoparticles ensembles in a highly dynamic magnetic field. Exploiting Maxwell's equations then allows for the determination of the entire MPI system equation \cite{Knopp2012r} which is discussed in the following.

\subsection{MPI system equation }
In general, the system equation in MPI is described by a linear integral equation in which the spatial SPIO distribution $c: \Omega \subset \R^3 \to \mathbb R$ is related to the measured time-varying voltage signal $u_\ell:\mathbb R\to\mathbb R$ by
\begin{equation} \label{eq:systemEquation}
	u_\ell(t) = \int_{\mathbb R^3} s_\ell(\bm{x},t) c(\bm{x}) \dt[\bm{x}],
\end{equation}
where $s_\ell:\mathbb R^3\times \mathbb R \to \mathbb R$ denotes the system function
\begin{equation}
s_\ell(\bm{x},t) = -\mu_0 \int_\R a_\ell(t-\tau)  \left(\bm{p}_\ell(\bm x)\right)^\tran\pnpt[\tau]{} \bm\Gamma_{\bar{\bm{m}}}(\tau,  \bm{H}(\bm{x},\cdot)) \dt[\tau].
\label{eq:systemfunc}
\end{equation}
The system function $s_\ell(\bm{x},t)$ itself is composed  of \mbox{$\bm{p}_\ell :\Omega \to\mathbb R^3$}, the $\ell$-th sensitivity profile  of the receive coil unit, the analog filter $a_\ell: \R \to \R$, and the time derivative  of the mean magnetic moment $\bm\Gamma_{\bar{\bm{m}}}:   \R \times C(\R,\R^3)\to\mathbb R^3$, which describes the mean magnetic moment of an SPIO as a function of the spatio-temporally changing magnetic field $\bm{H}:\mathbb R^3\times\mathbb R \to \mathbb R^3$ (which we assume to be at least continuous with respect to time). The notation $\bm{H}(\bm{x},\cdot)$ means that the function $\bm{H}$ is parameterized in the first argument and remains a function with respect to the second argument. Besides, $\mu_0$ is the vacuum permeability. 

In the context of the present work, we need to distinguish two cases for $\bm\Gamma_{\bar{\bm{m}}}$: 
\begin{itemize}
    \item Equilibrium models typically allow for a representation with respect to point evaluation in time of the applied field such that we can represent $\bm\Gamma_{\bar{\bm{m}}}$ as 
    \begin{equation}
      \bm\Gamma_{\bar{\bm{m}}}(t,  \bm{H}(\bm{x},\cdot))=\bar{\bm{m}}(\bm{H}(\bm{x},t))
      \label{eq:systemfunc_mbar_EQ}
    \end{equation}
    with $\bar{\bm{m}}:\R^3 \to \R^3$.
    \item Magnetization dynamics such as Brownian and Néel rotation require the solution of a differential equation. In this case, the solution depends on the entire history of the applied field. Here,  $\bm\Gamma_{\bar{\bm{m}}} $ already provides the most specific representation.
\end{itemize}

In this work, different mathematical models for modeling  $\bm\Gamma_{\bar{\bm{m}}}$ are studied in comparison to a fully measured system matrix. These are the FP model with Néel rotation, two equilibrium models without and with anisotropy, and a reduced equilibrium model with anisotropy based on the theoretical considerations for direct Chebyshev polynomial-based reconstruction in \cite{droigk2022direct}.
Common to all the approaches presented, with the exception of the reduced equilibrium model, is that they attempt to model the mean magnetic moment by the probability density function (PDF) $p(\bm{m},t,\beta \bm{H}(\bm{x},\cdot);\mathbb O)$  such that
\begin{equation}
	\bm\Gamma_{\bar{\bm{m}}}(t,  \bm{H}(\bm{x},\cdot))= \mo \int_{\mathbb S^2}  \bm m p(\bm{m},t,\beta \bm{H}(\bm{x},\cdot);\mathbb O) \dt[\bm m],
	\label{eq:MeanMagMomGeneral}
\end{equation}
where $\mathbb S^2$ denotes the surface of the unit sphere, $\mathbb O$ denotes the physical and model-dependent parameter sets, $m_0$ is the magnitude of the magnetic moment of one nanoparticle. The physical parameter $\beta=\tfrac{\mu_0\mo}{\kB \TP}$, where $\kB$ is the Boltzmann constant and  $\TP$ is the temperature of the SPIOs.
For the equilibrium case analogously to \eqref{eq:systemfunc_mbar_EQ} we can use the simplified representation 
\begin{equation}
\mbar{  \bm{H}(\bm{x},t);\mathbb O} = \mo \int_{\mathbb S^2}  \bm m p(\bm{m},\beta \bm{H}(\bm{x},t);\mathbb O) \dt[\bm m],
\label{eq:MeanMagMomEquilibrium}
\end{equation}
with respect to a PDF $p(\bm{m}, \bm{\xi};\mathbb O)$, $\bm{\xi}\in \R^3$, which itself has no time-dependence. Time-dependence is solely introduced via the point evaluation in time of the applied field $\bm H$. 
Note that all models and in particular the differential equation-based models are parametric in $\bm x$.
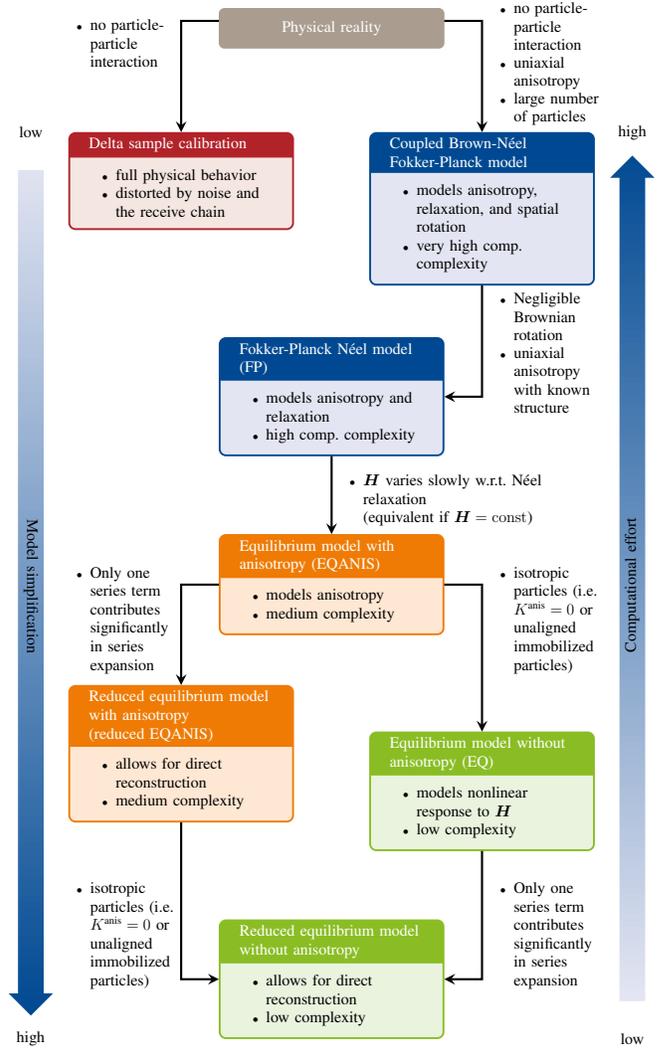
\begin{figure}[!t]
    \centering


\newcommand{\fontS}{\large}
\newtcolorbox{mybox_b}[1]
		{before={\vspace{-4pt}}, 
		colback = ibidarkBG, colframe = ibidark,  
		arc = 4pt, 
		title=#1,
		}
\newtcolorbox{mybox_o}[1]
		{before={\vspace{-4pt}}, 
		colback = ukesec2BG, colframe = ukesec2,  
		arc = 4pt, 
		title=#1,
		} 
\newtcolorbox{mybox_g}[1]
		{before={\vspace{-4pt}}, 
		colback = ukesec4BG, colframe = ukesec4,  
		arc = 4pt, 
		title=#1,
		}    
\newtcolorbox{mybox_r}[1]
		{before={\vspace{-4pt}}, 
		colback = ukesec5BG, colframe = ukesec5,  
		arc = 4pt, 
		title=#1,
		}      
\tikzstyle{boxnode} = [
	inner sep = -0.3pt, 
	text width = 6cm, 
 font = \fontS
	] 

\hspace*{0pt}
\begin{tikzpicture}[node distance=5cm and 8cm,on grid,scale=0.5,transform shape]
	
    \node (FPCoupled) [boxnode] {
    \begin{mybox_b}{Coupled Brown-Néel Fokker-Planck model}
		\begin{itemize}
			\item models anisotropy, relaxation, and spatial rotation
			\item very high comp. complexity
		\end{itemize}
	\end{mybox_b}                    
	};

	\node (FPNeel) [boxnode, below = of FPCoupled,xshift=-4cm] {
	\begin{mybox_b}{Fokker-Planck Néel model (FP)}
    	\begin{itemize}
    		\item models anisotropy and relaxation
    		\item high comp. complexity
    	\end{itemize}
 	\end{mybox_b}  
	};
 
	\node (EQWithAniso) [boxnode,below = of FPNeel] {
	\begin{mybox_o}{Equilibrium model with anisotropy (EQANIS)}
    	\begin{itemize}
    		\item models anisotropy 
    		\item medium complexity
    	\end{itemize}
    \end{mybox_o} 
	};
 
	\node (EQWithoutAniso) [boxnode,xshift=4cm,yshift=-0.5cm,below = of EQWithAniso] {
	\begin{mybox_g}{Equilibrium model without anisotropy (EQ)}
		\begin{itemize}
            \item models nonlinear response to  $\bm{H}$ 
		      \item low complexity
        \end{itemize}
    \end{mybox_g}  
	};
 
	\node (Chebyshev1) [boxnode,xshift=-4cm,yshift=0.5cm,below = of EQWithAniso] {
	\begin{mybox_o}{Reduced equilibrium model with anisotropy \\(reduced EQANIS)}
		\begin{itemize}
		      \item allows for direct reconstruction
		      \item medium complexity
	    \end{itemize}
    \end{mybox_o}       
	};
 
	\node (Chebyshev2) [boxnode,xshift=-4cm,below = of EQWithoutAniso] {
	\begin{mybox_g}{Reduced equilibrium model without anisotropy}
		\begin{itemize}
		      \item allows for direct reconstruction
		      \item low complexity
	    \end{itemize}
    \end{mybox_g}         
	};
	\path let \p1=(Chebyshev1), \p2=(FPCoupled) in node (FullPhysical) [boxnode] at (\x1/2+\x2/2,\y2+5cm) {
    \begin{tcolorbox}[colback = uke2, colframe = uke2]
        \center \color{white} \large Physical reality
    \end{tcolorbox}
    };
	
	\node (Measured) [boxnode,left = of FPCoupled.north, anchor=north] {
	\begin{mybox_r}{Delta sample calibration}
        \begin{itemize}
            \item full physical behavior 
		      \item distorted by noise and \\ the receive chain
	    \end{itemize}
    \end{mybox_r}       
	};

    \path let \p1=(Measured.west),\p2=(Measured.north) in node[font=\fontS] (leftArrowTop) at (\x1-1cm,\y2) {low};
    \path let \p1=(Measured.west),\p2=(Chebyshev2.south) in node[font=\fontS] (leftArrowBottom) at (\x1-1cm,\y2) {high};

    \path let \p1=(FPCoupled.east),\p2=(FPCoupled.north) in node[font=\fontS] (rightArrowTop) at (\x1+1cm,\y2) {high};
    \path let \p1=(EQWithoutAniso.east),\p2=(Chebyshev2.south) in node[font=\fontS] (rightArrowBottom) at (\x1+1cm,\y2) {low};


    \node(leftArrowMiddle)[fit=(leftArrowTop) (leftArrowBottom)]{};
    
    \node[single arrow,  minimum width=0.9cm, minimum height=22.5cm, draw=none, top color=ibidarkBG, bottom color=ibidark, rotate=0, line width=2pt,font=\fontS] at (leftArrowMiddle)[rotate=-90,align=center] {\textcolor{black}{Model simplification}};

    \node(rightArrowMiddle)[fit=(rightArrowTop) (rightArrowBottom)]{};
    
    \node[single arrow,  minimum width=0.9cm, minimum height=22.5cm, draw=none, top color=ibidark, bottom color=ibidarkBG, rotate=0, line width=2pt,font=\fontS] at (rightArrowMiddle)[rotate=90,align=center] {\textcolor{black}{Computational effort}};

	\draw[arrow](FullPhysical.east) -| node[right,text width=3cm,align=center,yshift=-1cm,font=\fontS] {
	\begin{itemize}
			\setlength{\itemsep}{-1pt}
        \item no particle-particle interaction
		\item uniaxial anisotropy
		\item large number of particles  
	\end{itemize}
	} (FPCoupled.north);

	\draw[arrow](FullPhysical.west) -| node[left,text width=3cm,align=left,yshift=-0.45cm,font=\fontS] {
	\begin{itemize}
			\setlength{\itemsep}{-1pt}
		\item no particle-particle interaction 
	\end{itemize}
	} (Measured.north);
	
		\draw[arrow](FPCoupled.south) |- node[right,text width=3cm,align=center,yshift=1.3cm,font=\fontS] {
	\begin{itemize}
			\setlength{\itemsep}{-1pt}
            \item\textcolor{black}{ Negligible Brownian rotation}
		\item \textcolor{black}{
                uniaxial anisotropy with known structure}
	\end{itemize}
	} (FPNeel.east);

	\draw[arrow](FPNeel) -- node[right,text width=7cm,align=left,font=\fontS] {
	\begin{itemize}
			\setlength{\itemsep}{-1pt}
		\item $\bm{H}$ varies slowly w.r.t. Néel relaxation \\ (equivalent if $\bm{H}=\mathrm{const}$)
	\end{itemize}
	} (EQWithAniso.north);
	\draw[arrow](EQWithAniso.east) -| node[right,text width=3cm,align=left,yshift=-0.8cm,font=\fontS] {
	\begin{itemize}
			\setlength{\itemsep}{-1pt}
		\item isotropic particles (i.e. $\Kanis = 0$ or unaligned immobilized particles) 
	\end{itemize}
	} (EQWithoutAniso.north);
	\draw[arrow](EQWithAniso.west) -| node[left,text width=3cm,align=center,yshift=-0.8cm,font=\fontS] {
	\begin{itemize}
			\setlength{\itemsep}{-1pt}
		\item Only one series term contributes significantly in series expansion 
	\end{itemize}
	} (Chebyshev1);
	
	\draw[arrow](EQWithoutAniso.south) |- node[right,text width=3cm,align=center,yshift=1.3cm,font=\fontS] {
	\begin{itemize}
			\setlength{\itemsep}{-1pt}
		\item Only one series term contributes significantly in series expansion 
	\end{itemize} 
	} (Chebyshev2.east);
	\draw[arrow](Chebyshev1.south) |- node[left,text width=3cm,align=left,yshift=1.3cm,font=\fontS] {
	\begin{itemize}
			\setlength{\itemsep}{-1pt}
		\item isotropic particles (i.e. $\Kanis = 0$ or unaligned immobilized particles)
	\end{itemize}
	} (Chebyshev2.west);

\end{tikzpicture}
    \caption{Overview of considered particle models and their underlying assumptions. The physical reality (top) can be modeled using a Delta-sample approach, assuming no particle interactions, i.e. a linear dependency of the particle response on the concentration. The same assumption is made in the coupled FP model; the FP model itself contains the assumption of a large ensemble of particles. In addition, a uniaxial anistropy is also assumed here. If the Brownian rotation is negligible and one has a known model for the uniaxial anistropy, then one has only pure Néel rotational dynamics of the magnetic moment. When additionally assuming instantaneous relaxation, one obtains the EQANIS model. This model can be reduced in two ways. First of all, one can assume that the Chebyshev series, which can be derived for 2D Lissajous excitation, is reduced to a single coefficient (reduced EQANIS model), see \cite{droigk2023adaption}. Second, one can assume isotropic particles leading to the EQ model. Both assumptions can be combined to the reduced EQ model, which was investigated for direct reconstruction in \cite{droigk2022direct}. }  \label{fig:ModelRelations}
\end{figure}

An overview of the considered models and the relation between the models is given in Fig.~\ref{fig:ModelRelations}.
As shown, all mathematical models presented are based on the assumption that the particle-particle interaction is negligible, one has reached a large particle number, and there is a uniaxial anisotropy in the particles. However, in \cite{Kluth2019} an approximate model for the system function of fluid particles was presented, which is heuristically motivated by the hypothesis that the static selection field causes a preferred easy axis orientation, which is used in the immobilized Néel relaxation FP model with space-dependent particle parameters.

In contrast to the presented models that directly model the PDF, stochastic differential equation models are also known in which the magnetic moment of each SPIO is directly modeled \cite{Weizenecker2012MicroMagneticSimulation,Shah2015,Graeser2016,neumann2020abitraryAnisotropies}.  The mean magnetic moment required for MPI is then estimated by averaging a certain number of simulated SPIOs in a volume of interest. Although this approach is much more ``natural'' for various physical effects, e.g., for arbitrary anisotropies \cite{neumann2020abitraryAnisotropies}, the drawback is that due to the stochastic nature of the simulation, a large number of particles must be simulated to achieve an accurate estimate of the mean magnetic moment, making the simulation very time-consuming. Note that in the particular setting of the present work, both approaches become equivalent in the large number of particle limit due to its relation via the Fokker-Planck equation. However, it should not go unmentioned that the same drawback of time-consuming simulation also applies to the Néel relaxation FP model, which makes the application of the proposed analytical equilibrium models attractive.

\subsection{Fully Data Driven (Delta Sample Calibration)}

Since proper modeling is still one of the key challenges in MPI, the most common method to determine the system function $s_\ell(\bm x,t)$ is based on time-consuming calibration measurements, where the system is effectively measured voxel by voxel. To this end, a small $\Delta$ sample, which is, for example, often of cuboid shape, is filled with tracer and shifted to positions $\bm x_n, n=1, \dots, N$, where $N$ is the total number of considered voxels. 
When assuming that the reference sample $\Delta$ is located at $\bm x_n$, the sample concentration $c_{\Delta, \bm x_n}(\bm x)$ can be modeled by $c_{\Delta, \bm x_n}(\bm x) = c_0 \Delta (\bm x - \bm x_n)$ where $c_0$ is the base concentration of the sample. $\Delta: \R^3 \rightarrow \{0,1\}$ is then the characteristic function of the reference sample. Inserting into \eqref{eq:systemEquation} leads to 
\begin{align*}
	u_{\ell, n}(t)& = \int_{\mathbb R^3} s_\ell(\bm x,t) c_0 \Delta (\bm x - \bm x_n) \dt[\bm x] \\ & \approx c_0 s_\ell(\bm x_n,t) \underset{=:V_\Delta}{\underbrace{\int_{\R^3} \Delta(\bm x )\dt[\bm x]}},
\end{align*}
which allows to directly estimate $s_\ell(\bm x_n,t)$ by dividing $u_{\ell, n}$ by the base concentration $c_0$ and the volume of the reference sample $V_\Delta$.
A full calibration procedure of the system matrix is typically designed in such a way that $\mathrm{supp}(c_{\Delta, \bm x_n})$, $n=1,\hdots,N$, are pairwise disjoint and that their union includes the desired field of view $\Omega_\text{SF}$.

The fully data driven model makes no physical assumptions other than the absence of particle-particle interactions resulting in the assumption of a linear relationship between concentration and voltage. It can thus be considered to be the most accurate model when the signal-to-noise ratio (SNR) of the calibration data is sufficiently high and that measurement noise can be neglected.

\subsection{Fokker-Planck N\'{e}el model}

When modeling the magnetization behavior of large ensembles of MNPs, Brownian and/or N\'{e}el rotation dynamics of the particle's magnetic moment are taken into account \cite{coffey2012langevin}. In order to obtain the mean magnetic moment of the MNP ensemble, solving the corresponding Fokker-Planck equation is one immediate opportunity to obtain the time-dependent probability density function $p$ required in \eqref{eq:MeanMagMomGeneral}. In line with existing works in the context of MPI \cite{Weizenecker2018,Kluth2018a,Kluth2019,albers2022simulating} we consider the following parabolic partial differential equation (PDE) on the sphere (Fokker-Planck equation for N\'{e}el rotation) with respect to $p: \mathbb{S}^2 \times \R \rightarrow \R$:
\begin{equation}
\label{eq:FP-general}
 \frac{\partial}{\partial t} p = \mathrm{div}_{\mathbb{S}^2}\left(\frac{1}{2\tau} \nabla_{\mathbb{S}^2} p \right) - \mathrm{div}_{\mathbb{S}^2}\left(\bm{b} p\right)
\end{equation}
where $\tau >0$ is the relaxation time constant and the (velocity) field $\bm{b}:\mathbb{S}^2 \times \R^3 \times \mathbb{S}^2 \rightarrow \R^3$ given by
\begin{align}
 &\bm{b}(\bm  m,\bm{H},\bm n) = \alpha_{\mathrm{N},1} \bm{H} \times \bm m + \alpha_{\mathrm{N},2} (\bm m\times \bm{H}) \times \bm m \notag \\ & \quad+ \alpha_{\mathrm{N},3} (\bm n^\tran \bm m) \bm n \times \bm m + \alpha_{\mathrm{N},4} (\bm n^\tran \bm m) (\bm m\times \bm n) \times\bm m \label{eq:convection}
\end{align}
with $\alpha_{\mathrm{N},i}\geq 0$, $i=1,\hdots,4$, being the physical constants for N\'{e}el rotation and $\bm n\in \mathbb{S}^2$ being the easy axis of the particles. 
Differentiation in terms of gradient $\nabla_{\mathbb{S}^2} $ and divergence $ \mathrm{div}_{\mathbb{S}^2}$ is considered with respect to the surface $\mathbb{S}^2$.
The numerical solution is obtained via the method of lines, i.e., by spatial discretization using, for example, spherical harmonics or functions defined on a triangular grid as basis functions, and an ODE system with respect to time is obtained. 
The ODE system is implemented and solved in Julia using the MNPDynamics toolbox \cite{albers2022simulating}.

\subsection{Equilibrium Model with Anisotropy}

Considering now the anisotropy effects to be taken into account in the Néel-FP model, i.e. according to the Stoner-Wohlfarth model \cite{Stoner1948},  the anisotropy is modeled using a uniaxial easy axis $\bm{n} \in \mathbb S^2$ and an anisotropy strength $\ak = \tfrac{\Vc \Kanis }{\kB \TP}$, where $\Kanis$ is the anisotropy constant and $\Vc$ is the core volume of an SPIO. In order to exploit approximate models for the fluid case \cite{Kluth2019}, we consider the parametric variants with respect to $\bm x$, i.e, \mbox{$\bm{n}:\mathbb R^3\to\mathbb S^2$} and \mbox{$\ak:\mathbb R^3\to\mathbb R$}. Therefore, the observable parameter set $\mathbb O(\bm x) = \{\ak(\bm x),\bm{n}(\bm x)\}$, where \mbox{$\bm x\in\mathbb R^3$} denotes the spatial position, is spatially variant. However, as it is solely parametric with respect to $\bm{x}$, for the sake of clarity, we omit the spatial variable $\bm x$ in $\mathbb O$ and $\bm{H}$ in the following. Thus, in thermodynamical equilibrium, the FP model in \eqref{eq:FP-general} for the PDF of the form in $\eqref{eq:MeanMagMomEquilibrium}$ has the solution \cite{albers2022immobilized,maass2022analytical} 
\begin{equation}
	p(\bm{m},\beta \bm{H};\mathbb O) = \frac{1}{\mathcal Z(\beta \bm{H};\mathbb O) }\eUnit^{ \beta \bm{H}^\tran \bm{m} +\ak (\bm{n}^\tran \bm{m})^2}
    \label{eq:BoltzmannAniso}
\end{equation}
with
\begin{equation}
\begin{aligned}
	\mathcal Z(\beta \bm{H};\mathbb O)  &= \int_{\mathbb{S}^2} \eUnit^{ \beta \bm{H}^\tran \bm{m} +\ak (\bm{n}^\tran \bm{m})^2} \dt[\bm{m}]\\
 &= \int_{\mathbb{S}^2} \eUnit^{ \beta (\bm{R}_{\bm{n}}^\tran\bm{H})^\tran \tilde{\bm{m}} +\ak \tilde{m}_3^2} \dt[\tilde{\bm{m}}],
 \end{aligned}
	\label{eq:partionAniso}
\end{equation}
where the rotation matrix $\bm{R}_{\bm{n}}\in\R^{3\times 3}$ is used in the second expression with $\bm{e}_3 = \bm{R}^\tran_{\bm{n}}\bm{n}$ and $\bm{e}_3\in\R^3$ being the third Euclidean unit vector.
It should be mentioned that for the PDF of the form of \eqref{eq:BoltzmannAniso}
\begin{equation*}
	\begin{aligned}
	\mbar{ \bm{H};\mathbb O} &= \mo \int_{\mathbb S^2}  \bm{m} p(\bm{m},\beta \bm{H};\mathbb O) \dt[\bm{m}] \\
    &= \mo \frac{\int_{\mathbb S^2}  \bm{m} \eUnit^{\beta \bm{H}^\tran \bm{m}+\ak(\bm{n}^\tran\bm{m})^2} \dt[\bm{m}]}{\mathcal Z(\beta \bm{H};\mathbb O)}\\
    & =\frac{\mo}{\beta} \frac{\int_{\mathbb S^2}  \nabla_{\bm{H}} (\eUnit^{\beta \bm{H}^\tran \bm{m}+\ak(\bm{n}^\tran\bm{m})^2}) \dt[\bm{m}]}{\mathcal Z(\beta \bm{H};\mathbb O)} \\ &\overset{\dagger}{=} \frac{\mo}{\beta} \frac{  \nabla_{\bm{H}} \mathcal Z(\beta \bm{H};\mathbb O)}{\mathcal Z(\beta \bm{H};\mathbb O)} = \frac{\mo}{\beta}\nabla_{\bm{H}} \ln\left(\mathcal Z(\beta \bm{H};\mathbb O)\right),
		\end{aligned}
\end{equation*}
holds, where the step marked with $\dagger$ is allowed as $\eUnit^{\beta \bm{H}^\tran \bm{m}+\ak(\bm{n}^\tran\bm{m})^2}$ is totally differentiable  in $\bm{H}$. The mean magnetic moment $\mbar{ \bm{H};\mathbb O}$ can be used in $\eqref{eq:systemfunc_mbar_EQ}$.

Throughout the article, the following is written for this model for the mean magnetic moment 
	\begin{equation}
	\begin{aligned}
	\mbar{ \bm{H};\mathbb O } 	&= \mo \mathscrbf{E}(\beta  \bm{H}; \mathbb O ) = {\mo} \frac{\nabla_{\bm{H}}\ln\left(\mathcal Z(\beta \bm{H};\mathbb O)\right)}{\beta},
		\end{aligned}
		\label{eq:MeanMagMomentAniso}
	\end{equation}
	where $\mathscrbf{E}(\beta  \bm{H}; \mathbb O ) = \tfrac{1}{\beta}\nabla_{\bm H}\ln\left(\mathcal Z(\beta \bm{H};\mathbb O)\right)$ fulfills the role of the Langevin function \cite{rahmer2009signal} and can be interpreted as a generalization of the Langevin function for anisotropic particles, where the uniaxial easy axis is aligned and immobilized.
\subsubsection{Equilibrium Model without Anisotropy}

The equilibrium model without anisotropy is only a special case of the model with anisotropy. 
If the particles are in thermodynamical equilibrium and particle anisotropies are neglected, i.e $\ak = 0$, and therefore the special choice of $\bm n\in \mathbb S^2$ no longer has any influence then $\isoParaSet=\{0,\bm{n}\}$ and the PDF related to the FP model in \eqref{eq:FP-general} is of the form  \eqref{eq:MeanMagMomEquilibrium}. Thus, \eqref{eq:BoltzmannAniso} becomes
\begin{equation}
	p(\bm{m},\beta \bm{H};\isoParaSet) = \frac{1}{\mathcal Z(\beta \bm{H};\isoParaSet) }\eUnit^{ \beta \bm{H}^\tran \bm{m} }
	\label{eq:BoltzmannIso}
\end{equation}
with $\bm H \in \R^3$ and the partition function
\begin{equation} 
\mathcal Z(\beta \bm{H};\isoParaSet) = 4\pi  \frac{\sinh(\beta \| \bm{H}\| )}{\beta \| \bm{H}\|} 
 \end{equation} 
such that $p(\bm{m},\beta \bm{H};\isoParaSet)$ is a PDF with respect to $\bm{m}$ (see also \cite{Kluth2018a} for a derivation).
The evaluation of \eqref{eq:MeanMagMomEquilibrium} with \eqref{eq:BoltzmannIso} yields
\begin{equation}
	\mbar{ \bm{H};\isoParaSet} = \mo \mathscr L (\beta \|\bm{H}\|_2) \frac{\bm{H}}{\|\bm{H}\|_2},
	\label{eq:MeanMagMomLang}
\end{equation}
where 
\begin{equation*}
\mathscr L (\xi) = \begin{cases}\coth(\xi)-\frac{1}{\xi} & \text{if }\xi\neq 0 \\
																0 &\text{if } \xi = 0
										\end{cases} 
\end{equation*}
is the Langevin function. 
Note that this $\mbar{ \bm{H};\isoParaSet}$ can be used in $\eqref{eq:systemfunc_mbar_EQ}$.

The resulting model is commonly referred to as the classical Langevin theory of paramagnetism and is used in MPI as a simplified attempt to model the MPI system function in \eqref{eq:systemfunc} since the very beginning of MPI research \cite{Weizenecker2007, rahmer2009signal}.

\subsubsection{Computational Aspects} \label{sec:compAspec}

The mean magnetic moment for the PDF in \eqref{eq:BoltzmannAniso} can no longer be solved in closed form. Therefore, the integrals in \eqref{eq:MeanMagMomEquilibrium} or \eqref{eq:partionAniso} can only be solved by representing them in spherical coordinates and then integrating them numerically \cite{cregg1999integral,albers2022immobilized} or by developing them in a series and truncating it \cite{cregg1999series,maass2022analytical}. We will follow the latter approach and in this work outline the derivations that have been skipped in the short proceedings paper \cite{maass2022analytical}.

The expression of \eqref{eq:partionAniso} in spherical coordinates requires the numerical solution of two iterated integrals. One of the iterated integrals can be solved, and a single integral expression is derived; see \cite{cregg1999integral} for details. Subsequently, the remaining spherical coordinate is modified by the substitution step for the presentation in this article. In a next step, we expand the functions in a Maclaurin series 
and, finally, the remaining integration is performed for each series term, individually. The last part of this subsection provides some information on the concrete implementation.

Using the substitution $\tilde{\bm{H}} = \bm{R}_{\bm{n}}^\tran \bm{H}$ with the help of the rotation matrix $\bm{R}_{\bm{n}}$ (see \eqref{eq:partionAniso}), according to \cite{maass2022analytical,cregg1999integral}, the integral expressions for $\mathcal Z(\bm{R}_{\bm{n}}\tilde{\vek H}; \mathbb O)$ 
and $\mbar{ \bm{R}_{\bm{n}} \tilde{\bm{H}};\mathbb O} $ are represented by
\begin{equation}
    \mbar{ \bm{R}_{\bm{n}} \tilde{\bm{H}};\mathbb O} = \mo \bm{R}_{\bm{n}} \frac{\left(z_j(\tilde{\vek H})\right)_{j=1}^3}{\mathcal Z(\bm{R}_{\bm{n}}\tilde{\vek H};\mathbb O)}
\end{equation}
with
\begin{align}
 \mathcal Z (\bm{R}_{\bm{n}}\tilde{\vek H};\mathbb O) &= 4 \pi \int_{0}^{1} I_0\left(\beta |\tilde{\vek H}|_{12}\sqrt{ 1- x^2} \right) \label{eq:IntExpressZ} \\ 
 &\hspace{3.3cm}\times\cosh(\beta \tilde{H}_3 x) \eUnit^{\ak x^2  }\dt[x], \nonumber
\end{align}
\begin{align}
 z_3(\tilde{\vek H}) &= 4 \pi \int_{0}^{1} x I_0\left(\beta |\tilde{\vek H}|_{12} \sqrt{ 1- x^2} \right) \label{eq:IntExpressZ3} \\ & \hspace{4cm}\times\sinh(\beta \tilde{H}_3 x) \eUnit^{\ak x^2  }\dt[x], \nonumber
 \end{align}
 \begin{align}
z_{i}(\tilde{\vek H}) &=  4\pi \frac{ \tilde{H}_i  }{|\tilde{\vek H}|_{12}} \int_{0}^{1}   \sqrt{1- x^2}  I_1\left(\beta |\tilde{\vek{H}}|_{12} \sqrt{1- x^2} \right) \label{eq:IntExpressZ12}\\ &\hspace{4cm}\times\cosh(\beta \tilde{H}_3 x) \eUnit^{\ak x^2  }\dt[x] \nonumber
\end{align}
using $|\tilde{\vek{H}}|_{12} =  \sqrt{\tilde{H}_1^2+\tilde{H}_2^2}$, $\cosh(\xi) =\tfrac{\eUnit^{\xi}+\eUnit^{-\xi}}{2}$, $\sinh(\xi) =\tfrac{\eUnit^{\xi}-\eUnit^{-\xi}}{2}$, and $i\in\{1,2\}$.
Note that $\tilde{H}_3 = \bm{n}^\tran \bm{H}$ and $|\tilde{\bm{H}}|_{1,2} = \|\bm{H}-\tilde{H}_3 \bm{n}\|_2$ are the colinear and orthogonal contributions with respect to the easy axis $\bm{n}$ of the magnetic field $\bm{H}$.

\begin{figure}[t]
\begin{tlemma}[label=lem:seriesExansion]{Series Expansion}{}
The integral representations  \eqref{eq:IntExpressZ}, \eqref{eq:IntExpressZ3}, and \eqref{eq:IntExpressZ12} can be expressed by 
\begin{align}
    &\mathcal{Z}(\bm{R}_{\bm{n}}\tilde{\bm H};\mathbb O) = (2\pi)^\frac{3}{2} \sum_{\ell=0}^\infty \Bigg[(2\ak)^\ell L_\ell^{(-\frac{1}{2})}\mathopen{}\left(-\frac{\beta^2 \tilde{H}_3^2}{4\ak}\mathclose{}\right) \nonumber\\
    & \hspace{5cm}\times\frac{I_{\ell+\frac{1}{2}}(\beta|\tilde{\vek{H}}|_{12})}{(\beta|\tilde{\vek{H}}|_{12})^{\ell +\frac{1}{2}}} \Bigg] \label{eq:seriesZ}\\
    &z_3(\tilde{\bm H}) =  (2\pi)^\frac{3}{2} \beta \tilde{H}_3 \sum_{\ell=0}^\infty\Bigg[ (2\ak)^\ell  L_\ell^{(\frac{1}{2})}\mathopen{}\left(-\frac{\beta^2 \tilde{H}_3^2}{4\ak}\mathclose{}\right) \nonumber\\ 
    &\hspace{5cm}\times\frac{I_{\ell+\frac{3}{2}}(\beta|\tilde{\vek{H}}|_{12})}{(\beta|\tilde{\vek{H}}|_{12})^{\ell +\frac{3}{2}}}\Bigg]\label{eq:seriesz3}\\
    &z_i(\tilde{\bm H}) =  (2\pi)^\frac{3}{2} \beta \tilde{H}_i \sum_{\ell=0}^\infty\Bigg[ (2\ak)^\ell L_\ell^{(-\frac{1}{2})}\mathopen{}\left(-\frac{\beta^2 \tilde{H}_3^2}{4\ak}\mathclose{}\right) \nonumber\\
    &\hspace{5cm}\times\frac{I_{\ell+\frac{3}{2}}(\beta|\tilde{\vek{H}}|_{12})}{(\beta|\tilde{\vek{H}}|_{12})^{\ell +\frac{3}{2}}}\Bigg]\label{eq:seriesz12},
\end{align}
where \mbox{$L_n^{(\alpha)}:\R\to\R$} denotes the generalized Laguerre polynomials with order $\alpha>-1$  and degree $n\in\N_0$ and \mbox{$I_{\nu}:\C\to\C$} denotes the modified Bessel function of first kind with order $\nu\in\C$.
\end{tlemma}
\end{figure}

Lemma \ref{lem:seriesExansion} shows, that \eqref{eq:IntExpressZ}, \eqref{eq:IntExpressZ3}, and \eqref{eq:IntExpressZ12} can be instead expressed by series expressions. The proof will be given in the supplementary material A.
The advantage of the series expression is that the numerical integration can be mitigated and instead a summation can be performed. 
 We expect the summands for large orders $\ell$ to decrease very quickly, since the modified Bessel functions have the following asymptotic expansion \mbox{$\tfrac{I_\nu(z)}{z^\nu}\sim \frac{\eUnit^\nu}{2^\nu \nu^\nu \sqrt{2\pi\nu}}$} \cite[§9.3.1]{Abramowitz1964} and for the generalized Laguerre polynomials the following asymptotic expansion applies \mbox{$L^{(\alpha)}_n(-\xi)\sim C(\alpha,\xi) n^{\alpha/2-1/4} \eUnit^{2\sqrt{n \xi}}$} with $C(\alpha,\xi)$ constant for fixed $\xi>0$ and $\alpha>-1$ \cite{Borwein2008}. 
Using the two expansions, the convergence-dominant terms of the series, e.g. for \eqref{eq:seriesZ}, are given by $\frac{\ak^\ell\eUnit^{\ell+1/2}}{(\ell+1/2)^{\ell+1/2}}$.  The term $\tfrac{1}{(\ell+1/2)^{\ell+1/2}}$ has the strongest effect and a super-exponential decay, so that a rapid decay is to be expected for sufficiently large $\ell$.
Thus, for the actual implementation, the series in \eqref{eq:seriesZ}, \eqref{eq:seriesz3}, and \eqref{eq:seriesz12} were truncated to $L$ series terms so that the summation is performed from $\ell=0$ to $\ell=L-1$. As the modified Bessel functions of fractional order and generalized Laguerre polynomials  can have large values potentially leading to numerical overflow, logarithmic scales are used as far as possible in the current implementation. For the implementation of generalized Laguerre polynomials the recurrent relation for orthogonal polynomials is used.  The actual implementation is provided in the MNPDynamics toolbox in the Julia programming language \cite{Albers2020IWMPI}.  

\subsection{Reduced Equilibrium Model with Anisotropy} \label{sec:redEq}

Since for Lissajous MPI sampling trajectories, a relation of the Fourier  coefficients $s_{\ell k}(\bm x)$ (\mbox{$k\in\mathbb Z$}) of the system function $s_\ell(\bm x,t)$ in \eqref{eq:systemfunc} to tensor products of weighted Chebyshev polynomials has long been observed \cite{rahmer2009signal, rahmer2012analysis} and was shown in \cite{maass2020representation} for the isotropic model in \eqref{eq:MeanMagMomLang}, it seems reasonable to formulate this model also for the anisotropic case in \eqref{eq:MeanMagMomentAniso}, which has only been investigated numerically so far \cite{maass2023system}.
The reduced equilibrium model without anisotropy is the underlying model of the Chebyshev polynomial-based reconstruction method in~\cite{droigk2022direct}, but it can be formulated for the equilibrium model with anisotropy as well~\cite{droigk2023adaption}, which made it a model worth investigating within this work. 

However, the introduction of the reduced model requires additional assumptions about the applied field   $\bm{H}(\bm x,t)$.  	The applied field $\bm{H}(\bm x,t) = \HS(\bm x)+\HD(t)$ is a superposition of two idealized magnetic fields, the selection field \mbox{$\HS(\bm x) = \bm G \bm x$}, which is a linear gradient field with $\bm G\in\mathbb R^{3\times 3}$ being an invertible diagonal matrix with $\bm G= \operatorname{diag}(G_x,G_y,G_z)$, and the drive-field \mbox{$\HD:\mathbb R\to\mathbb R^3$}, which is a time-varying and $\TD$-periodic magnetic field. Let the drive-field be defined by $\HD(t)=(A_x\sin(2\pi f_x t +\varphi_x),A_y\sin(2\pi f_y t +\varphi_y),0)^\tran$ with $A_x,A_y\in\R\backslash\{0\}$, $\varphi_x,\varphi_y\in\R$, $f_x=\frac{\fB}{\NB+1}$, $f_y=\frac{\fB}{\NB}$, $\fB>0$ the base frequency of the MPI scanner and $\NB\geq 1$ the frequency divider. In addition, the duration of one period of excitation is $\TD = \tfrac{\NB(\NB+1)}{\fB}$. This results in an MPI scanner with a field-free point (FFP) that travels along a Lissajous trajectory.  
 For simplicity, let also $\bm{p}_\ell(\bm{x}) = \bm{\rho}_\ell$ with $\bm{\rho}_\ell \in\mathbb R^3$ for any $\bm x \in \R^3$.
	
In \cite[Eq. (8)]{maass2023system} it was numerically shown, but not formally proven, that the Fourier series components  $s_{\ell k}(\bm{x})$ of the system function $s_\ell(\bm{x},t)$ for the equilibrium model of SPIOs 
with anisotropy and with a two-dimensional Lissajous FFP-trajectory can be expressed by 
\begin{equation}
	\begin{aligned}
		s_{\ell k}(\bm{x}) = \iUnit \omega_k  \bm{M}_{\ell k}^\tran \!\!\int_{\mathbb R^3} \left[\frac{\partial^2}{\partial z_1 \partial z_2}\mathscrbf{E}(\beta \bm{G} \bm z; \mathbb O)\right]_{\scriptscriptstyle \bm z = \bm x- \bm y} P^{(2)}_k(\bm{y}) \dt[\bm{y}],
	\end{aligned}
	\label{eq:sysFunc2DCheb}
\end{equation}
where $P^{(2)}_k(\bm \xi)$ is related to a series of two-dimensional tensor products of weighted Chebyshev polynomials of second kind with $k\in\mathbb Z$,  $\omega_k = \frac{2\pi k}{\TD}$, and $\bm{M}_{\ell k} =-\hat{a}_\ell(\omega_k)\mu_0 \mo \bm{\rho}_\ell$. Besides, $\hat{a}_\ell:\R\to\C$ denotes the Fourier transform of the analog filter in \eqref{eq:systemfunc}, commonly referred to as the transfer function.

 According to \cite{maass2020representation}, using the specific drive-field sequence, we obtain for the series of two-dimensional tensor products of weighted Chebyshev polynomials of second kind:
\begin{equation}
	\begin{aligned}
     P^{(2)}_k(\bm{x}) &= \frac{\sgn(\tfrac{A_x}{G_x})\sgn(\tfrac{A_y}{G_y})}{\pi^2} \delta_0(x_3) \\
      & \quad \times \sum_{\lambda\in\Z}\iUnit^{\lambda} \eUnit^{\iUnit\theta_{k}(\lambda)}V_{n_k(\lambda)}\!\left(\frac{G_x}{A_x}x_1\right)V_{m_k(\lambda)}\!\left(\frac{G_y}{A_y}x_2\right),
\end{aligned}
\end{equation}
where {$n_k(\lambda) = -k+\lambda (\NB+1)$}, $m_k(\lambda) = k-\lambda\NB$, {$\theta_{k}(\lambda)=n_k(\lambda)\varphi_x +m_k(\lambda)\varphi_y $}, and $\delta_0(\xi)$ denotes the delta distribution and is used to express \eqref{eq:sysFunc2DCheb} as a three-dimensional convolution. The function
\begin{equation}
    V_n(\xi) = 
    \begin{cases} 
    \rect\!\left( \frac{\xi}{2}\right)\left(-\frac{U_{|n|-1}(\xi)\sqrt{1-\xi^2}}{|n|}\right),& \text{if } |n|>0 \\
    \frac{\pi}{2}\sgn(\xi+1) -\rect\!\left( \frac{\xi}{2}\right)\arccos(\xi),& \text{if } |n|=0, \\
    \end{cases}
\end{equation}
is for $n\in\N$ composed of Chebyshev polynomials of the second kind \mbox{$U_{m}:\R \to\R$} with the degree  $m\in\N_0$ and the corresponding weighting function $\sqrt{1-\xi^2}$.

The reduced equilibrium model truncates the series of Chebyshev polynomials in $P^{(2)}_k\! (\bm \xi)$ to a single summand. 
For the derivation in~\cite{droigk2022direct}, the representation of the system function in~\eqref{eq:sysFunc2DCheb} was used.
For this, let $n_k^*=n_k(\lambda_k^*), m_k^*=m_k(\lambda_k^*)$ 
denote the choice of the orders of the Chebyshev polynomials of second kind and $\theta_k^* = \theta_k(\lambda_k^*)$ denotes the phase that depends on the drive-field. 
This can be specified 
with $\lambda_k^* = \operatorname{round}\!\left(\frac{2\NB k}{2\NB^2+2\NB+1}\right)$ for a two-dimensional Lissajous-like excitation~\cite{droigk2022direct}.
 
Let 
\begin{equation}
	\begin{aligned}
    S_k^{(2)}(\bm x; \lambda^*_k) &= \frac{\sgn(\tfrac{A_x}{G_x})\sgn(\tfrac{A_y}{G_y}) \iUnit^{\lambda_k^*}\eUnit^{\iUnit\theta_k^*}}{\pi^2} \delta_0(x_3) \\ &
    \hspace{1cm} \times V_{n^*_k}\!\left(\frac{G_x}{A_x}x_1\right)V_{m_k^*}\!\left(\frac{G_y}{A_y}x_2\right),
\end{aligned}
\end{equation}
denote the specific summand of the series $P_k^{(2)}(\bm x)$ that is specified by the index $ \lambda^*_k$. 
Then, the Fourier series components of the system function for the reduced equilibrium model with anisotropy can be formulated analogously to~\eqref{eq:sysFunc2DCheb} as
\begin{equation}
	\begin{aligned}
		s_{\ell k}(\bm{x}) &\approx \iUnit \omega_k  \bm{M}_{\ell k}^\tran\!\!\int_{\mathbb R^3} \Bigl[\frac{\partial^2}{\partial z_1 \partial z_2}\mathscrbf{E}(\beta \bm{G} \bm z; \mathbb O)\Bigr]_{\scriptscriptstyle \bm z = \bm x- \bm y}\hspace{-0.4cm} S^{(2)}_k(\bm{y}, \lambda^*_k) \dt[\bm{y}].
	\end{aligned}
	\label{eq:sysFuncCompAnisoReduced}
\end{equation}

\subsection{Models for the anisotropy}
 
For modeling the anisotropy, we distinguish two cases. In the first case, the particles have been immobilized while being subjected to a magnetic field. In turn, the particles have an oriented easy axis and a constant anisotropy value
\begin{equation}
    \begin{aligned}
        \bm n(\bm x) &= \tilde{\bm n}\\
        \ak(\bm x) &= \frac{\beta}{\mu_0 \MS} \Kanis
    \end{aligned}
\end{equation}
with $\tilde{\bm n}\in\mathbb S^2$, $\Kanis\in\R$, and $\MS=\tfrac{\mo}{\Vc}$. 

For fluid particles and the special case of Lissajous sequences, we consider a slightly extended version of the heuristic model from \cite[B3]{Kluth2019} with a spatially varying anisotropy parameter set $\mathbb O(\bm x)$, which is chosen to be
\begin{equation}
    \begin{aligned}
        \bm n(\bm x) &= \frac{\bm H^S(\bm x)}{\|\bm H^S(\bm x)\|_2}\\
        \ak(\bm x) &= \frac{\beta}{\mu_0 \MS} \Kanismax 
        \left(\frac{\|\bm H^S(\bm x)\|_2}{h} \right)^q,
    \end{aligned}
\end{equation}
where $h=\max_{\bm x \in \Omega_\mathrm{SF}}\|\bm H^S(\bm x)\|_2$ is the maximum gradient strength at the scan-field boundaries and $\Kanismax\in\mathbb R$ is the anisotropy parameter. 
In this work, the new parameter \mbox{$q \in \mathbb{R}_+$} is introduced, which allows us to model a non-linearity in such a way that the anisotropy can change more rapidly in outer regions.
The motivation for this model is that the effective anisotropy constant $\ak(\bm x)$ includes a weighting that corresponds to the length of the mean of all easy axes over an entire sequence cycle. In the center, the mean would be zero, while it would have high value in outer positions, where the easy axes have a preferred direction. In \cite[B3]{Kluth2019} the value was $q=1$, but we recently observed that this value leads to system functions that are non-differentiable at position $\bm x = \bm 0$.

\section{Materials and Methods}

The different models are evaluated by simulating and comparing typical MPI system matrices using complex 2D Lissajous type MPI sequences.
In addition, all models are compared to measured data, both at the level of system matrices and after reconstruction of phantom data.

\subsection{Experimental Data}

Experimental data was acquired with the preclinical MPI scanner from Bruker (Ettlingen, Germany). All experiments were performed using a 2D sequence within the $xy$-plane (horizontal) of the scanner. The gradient is set to $-1 ~\text{T}/\text{m}/\mu_0$ in the $x$- and $y$-directions and $2 ~\text{T}/\text{m}/\mu_0$ in $z$-direction. The drive field frequencies are given by $f_x = 2.5~\text{MHz} / 102$ and $f_y = 2.5~\text{MHz} / 96$ while the drive field amplitudes are set to a nominal value of $A_x, A_y = 12~\text{mT}/\mu_0$ at the scanner console. 
The resulting Lissajous sampling trajectory has a frequency ratio of $\tfrac{f_y}{f_x}=\frac{17}{16}$. The tracer perimag (micromod, Rostock, Germany) was used for all experiments.

We performed two different studies, one of which was already published in \cite{albers2022modeling,moddel2021estimating}. This first dataset was measured with immobilized particles, with aligned easy axes. In this paper we only consider the case where the easy axis is rotated by $45^\circ$ in the $xy$-plane (see Fig.~\ref{fig:SMVisuImmobilizedOrientated}).  A system matrix was measured by applying offset fields to a static cylindrical delta sample in the center (20\,\textmu L perimag at a concentration of 89\,mmol(Fe)$^{-1}$ and 21\,\textmu L sodium alginate powder for immobilization). Data was collected for $11 \times 11$ offset values ranging from $-14.35~\text{mT}/\mu_0$ to $14.35~\text{mT}/\mu_0$ in $x$-direction and $-14.71~\text{mT}/\mu_0$ to $14.71~\text{mT}/\mu_0$ in $y$-direction. 

The second newly acquired dataset was measured with fluid perimag with a concentration of $10\,\text{mg}_{\text{Fe}}\text{mL}^{-1}$. In this case, the system matrix was measured by shifting the $\Delta$-sample to $17 \times 15$ positions covering a field-of-view (FOV) of $34 \times 30$\,mm$^2$. 
In addition to the system matrix, we measured 6 different phantoms to study the impact of modeling errors on the result after image reconstruction. The snake phantom consists of 5 cubic rods (2.5\,mm $\times$ 2.5\,mm cross section) of lengths 20\,mm, 17.5\,mm, 15\,mm, 8.75\,mm, 5\,mm that are arranged as a snake (see Fig.~\ref{fig:RecoFluid}). Three resolution phantoms were created using two rods of length 20\,mm and 17.5\,mm that are placed in parallel with three different distances (3\,mm, 5\,mm, and 7\,mm). The 5th phantom consists of a shape with a cone at the lower part and a sphere at the upper part resembling ice cream in a waffle. The 6th phantom is the $\Delta$-sample positioned 6\,mm apart from the center in the $x$ and the $y$ directions.

All measurements were background corrected using the method described in \cite{knopp2019correction}.  

\subsection{Model Parameters}

For simplicity, we consider a monodisperse particle size and anisotropy only. This case is often sufficiently accurate and reduces the parameter space substantially. The resulting particle parameters $D$, $K^\text{anis}$/$K^\text{anis}_\text{max}$, and $q$ were manually optimized by visual inspection of the resulting simulated system matrices. In the case of the fluid data, we also inspected the resulting model-based reconstruction results as a criterion. 

In addition to the particle model parameters, one also needs to model the transfer function of the receive path. To this end, we use the established data driven method that fits the modeled system matrices to a measured system matrix \cite{knopp2009model} using all positions.

\subsection{Image Reconstruction}

After discretization of the continuous imaging equation \eqref{eq:systemEquation} and applying a discrete Fourier transform, a linear system of equations
\begin{equation}
    \bm{S} \bm{c} = \bm{u}
\end{equation}
arises, where $\bm{S} \in \mathbb{C}^{M\times N}$ is the system matrix, $\bm{u} \in \mathbb{C}^{M}$ contains the measured voltages of all receive channels and $\bm{c} \in \mathbb{R}^{N}_+$ is the unknown particle concentration vector. As commonly done, we consider the measured voltage signals in frequency domain. Somewhat uncommonly for MPI reconstructions, we do not apply any frequency selection to the data.

For image reconstruction we solve the weighted and regularized least-squares problem 
\begin{equation} \label{eq:recoapproach}
   \underset{\bm{c}\geq \bm{0}}{\text{min}} \left(\Vert \bm{W} ( \bm{S} \bm{c} - \bm{u}) \Vert_2^2 + \lambda \Vert \bm{c} \Vert_2^2 \right).
\end{equation}
The weighting matrix $\bm{W}$ is chosen such that the noise is whitened according to a diagonal covariance matrix \cite{Kluth2019numerical}. The optimization problem \eqref{eq:recoapproach} is solved using the iterative Kaczmarz algorithm \cite{Knopp2010PhysMedBio}. We use a high number of 100 iterations such that convergence is reached. The regularization parameter $\lambda$ is chosen in a relative way \cite{Knopp2010PhysMedBio} ($\lambda = \lambda_0 \lambda_r$ with $\lambda_0 = \Vert \bm{W}\bm{S}\Vert_F^2/N$ ) and set $\lambda_r = 0.1$ in all cases.

\subsection{Software}

\begin{figure*}
    \centering
    \includegraphics[width=1.0\textwidth]{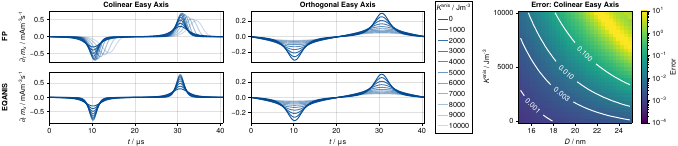}
    \caption{Comparison of the FP model and the EQANIS model based on a 1D simulation with a colinear (left) and an orthogonal (middle) easy axis with respect to the excitation direction. Shown are simulations for $D=20\,\textrm{nm}$ and $K^\text{anis}\in[0\,\textrm{Jm}^{-3}, 10\,000\,\textrm{Jm}^{-3}]$. On the right, the error $\varepsilon^\text{TD}$ is shown for the colinear easy axis, $D \in [15\,\textrm{nm}, 25\,\textrm{nm}]$ and the same anisotropy interval between the FP model ($\bar{\bm{m}}^\text{ref}$) and the EQANIS model ($\bar{\bm{m}}^\text{approx}$).}
    \label{fig:Accuracy1D}
\end{figure*}

The modeling parts of this paper have been integrated into the software package \textit{MNPDynamics.jl}, which is the Julia version of the Matlab reference implementation \cite{Albers2020IWMPI}. 
The code reproducing all results can be accessed under \url{https://github.com/IBIResearch/EquilibriumModelWithAnisotropy}. All data is automatically downloaded from Zenodo and can also be directly reached at \cite{knopp_2024_10646064}.

\subsection{Distance Measures}

For the comparisons of the models, we first consider the time signals $\bar{\bm{m}}^\text{ref}(\bm x,t)$ and $\bar{\bm{m}}^\text{approx}(\bm x,t)$, and calculate a global  error
\begin{align}
    	&\varepsilon^\text{TD}(\bar{\bm{m}}^\text{ref},\bar{\bm{m}}^\text{approx}) = \nonumber\\ &\quad \underset{\bm{x}}{\text{max}} \frac{\frac{1}{T} \int_{0}^{T} | \frac{\partial}{\partial t} \bar{\bm{m}}^\text{ref}(\bm{x},t) - \frac{\partial}{\partial t} \bar{\bm{m}}^\text{approx}(\bm{x},t) |  \textrm{d}t }{ \Vert  \frac{\partial}{\partial t} \bar{\bm{m}}^\text{ref}(\bm{x}, \cdot)  \Vert_\infty } \label{eq:errTD}.
\end{align}
The error is calculated over all spatial positions $\bm{x}$ where each one corresponds to a certain offset field.

For the measurements, we consider the error on the system matrix level in frequency space.
Let $\bm{s}_k^\text{ref} \in \mathbb{C}^N$ be the row of the measured reference system matrix at frequency index $k$ and $\bm{s}_k^\text{approx}\in \mathbb{C}^N$ be the corresponding approximating one. Then we calculate
\begin{align*}
    	&\varepsilon^\text{SM}(\bm{s}^\text{ref}_k, \bm{s}_k^\text{approx}) =  \frac{\Vert \bm{s}_k^\text{ref}- \bm{s}_k^\text{approx}\Vert_2}{ \sqrt{N} \Vert  \bm{s}_k^\text{ref} \Vert_\infty}.
\end{align*}
The frequency index $k$ here is chosen based on the mixing factors $\kappa_x,\kappa_y \in \mathbb{Z}$
with
$$k(\kappa_x,\kappa_y) = \kappa_x \NB + \kappa_y (\NB + 1)$$
for selected mixing orders outlined in the results. Here, \mbox{$\NB\in\N$} is the frequency divider (see Sec. \ref{sec:redEq}).

\section{Results}

\subsection{Accuracy}

In the first study, we compare the models directly on the level of the calculated magnetic moments. To this end, the particle core diameter is chosen in the range of $[15\,\textrm{nm}, 25\,\textrm{nm}]$ and the anisotropy is chosen in the range of  $[0\,\textrm{Jm}^{-3}, 10\,000\,\textrm{Jm}^{-3}]$. Both ranges enclose typical values that can be found in the model-based MPI literature. We note that the combination $D=25\,\textrm{nm}$ and $K^\text{anis} = 10\,000\,\textrm{Jm}^{-3}$ is already an edge case, in which the particles will exhibit stronger relaxation. We then perform simulations considering a 1D excitation in $x$-direction with 12 mT$/\mu_0$ using the excitation frequency $f_x$ from the Bruker scanner.  The resulting derivative $\frac{\partial}{\partial t} m_x$ for the central position is shown in Fig.~\ref{fig:Accuracy1D} for the FP model and the EQANIS model with a particle core diameter of 20~nm. Two cases are considered: The case that the easy axis is colinear and the case that the easy axis is orthogonal to the excitation direction. The simulated signals show the typical two peaks that occur when the vector field changes its direction. When increasing the anisotropy starting from $K^\text{anis} = 0\,\textrm{Jm}^{-3}$ to about $K^\text{anis} = 4000\,\textrm{Jm}^{-3}$, one can see that the peaks get higher and sharper. For these anisotropies, the EQANIS model and the FP model look basically the same. For larger anisotropies, relaxation can be observed for the FP model as the peaks are shifted in time. Since the EQANIS model depends only on the applied field at the current time point, it cannot express any type of relaxation leading to large deviations in those cases. No relaxation can be observed for the orthogonal easy axis and in turn both models look qualitatively the same.

For a quantitative error analysis, we calculate the error $\varepsilon^\text{TD}$ between the FP-model $\bar{\bm{m}}^\text{ref}$ and the EQANIS model $\bar{\bm{m}}^\text{approx}$ in dependence of the particle core diameter and the anisotropy for the colinear easy axis. Here, we consider a unit gradient field and all positions corresponding to  offset fields in the range of $[0\, \text{mT}/\mu_0, 12\, \text{mT}/\mu_0]$. Thus, the maximum error over all positions is considered (see \eqref{eq:errTD}). The results are shown in the  heatmap of Fig.~\ref{fig:Accuracy1D} on the right. One can see that the error increases both when increasing the diameter and when increasing the anisotropy. Several contour lines allow the parameter space to be separated into two regions where the error is either lower or higher than a predefined error. This can be used to predict how well certain particle parameters will be modeled by the EQANIS model compared to the FP model.

\begin{figure}
    \centering
    \includegraphics[width=1.0\columnwidth]{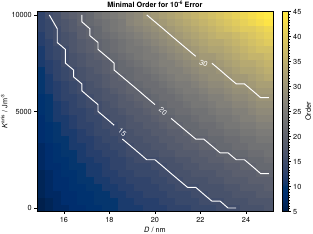}
    \caption{Minimum order $O$ that is required to reach an error of less than $10^{-6}$ in dependence of the particle diameter $D$ and the anisotropy $K^\text{anis}$ for a 1D simulation with $f_x = 2.5~\text{MHz} / 102$ and $A_x = 12~\text{mT}/\mu_0$. The reference signal was simulated with $L=200$.}
    \label{fig:Order}
\end{figure}

Within the previous simulations, the series truncation index $L$ within the EQANIS series implementation was set to a high value of $L=200$ (see Sec. \ref{sec:compAspec}), for which we assumed that the truncation error is negligibly small. To investigate influence of the truncation index we next repeat the simulation and calculate for each $D$ and $K^\text{anis}$ that truncation index $L^\text{min}(D, K^\text{anis})$ for which the error is smaller than $10^{-6}$. This value is chosen so that the truncation error is much smaller than the modeling error investigated before. The results are shown in Fig.~\ref{fig:Order}. One can see that the required number of summands also increases with  the diameter and the anisotropy. The maximum required truncation index is $L = 45$, which backs up our assumption that the series can be truncated at a rather small index. We use this value in all simulations performed next.

\begin{figure*}
    \centering
    \includegraphics[width=1.0\textwidth]{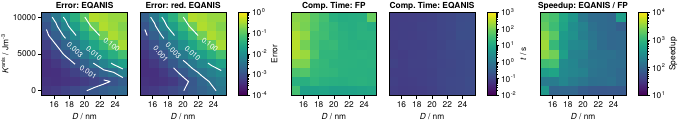}
    \caption{Error $\varepsilon^\text{TD}$ of the EQANIS and the reduced EQANIS model compared to the FP model considering 2D system matrix simulations with a static easy axis directed in $45^\circ$ to the $x$-axis in positive direction. The maximum error over all offset fields is calculated in dependence of the particle core diameter $D\in[15\,\textrm{nm},25\,\textrm{nm}]$ and the anisotropy $K^\text{anis}\in[0\,\textrm{Jm}^{-3}, 10\,000\,\textrm{Jm}^{-3}]$. The right part of the figure shows the computation time for the simulations using the Fokker-Planck and the EQANIS model.
    }
    \label{fig:Accuracy2D}
\end{figure*}

We then switch to the 2D excitation and calculate system matrices corresponding to the immobilized experimental dataset with a $45^\circ$ easy axis $\bm n$ and again calculate the error $\varepsilon^\text{TD}$ between the reference FP model and the EQANIS model as well as  the reduced EQANIS model. The results are shown in the left part of Fig.~\ref{fig:Accuracy2D}. One can see that the error map looks very similar, which is to be expected since it has been observed for the isotropic equilibrium model in previous works that the series in \eqref{eq:sysFunc2DCheb} is only dominated by its largest series term, which leads to the reduction strategy in \eqref{eq:sysFuncCompAnisoReduced} \cite{maass2020novel,droigk2022direct}.

\subsection{Computation Time}

We next take a look at the performance of the calculation of the system matrices, which is shown in the right part of Fig.~\ref{fig:Accuracy2D} as a heatmap. All calculations were performed using multi-threading with 64 threads. We first consider the computation time of the FP solver. As can be seen, the computation time is not constant but differs when varying the particle core diameter and the anisotropy. This is because the numerical solver uses an adaptive step width and thus might need more or less steps to reach a certain predefined accuracy, which was set to $\varepsilon^\text{rel} = 2\,\cdot\,10^{-4}$ and $\varepsilon^\text{abs} = 1\,\cdot\,10^{-6}$. There are certain combinations of diameter and anisotropy ($D=15\,\textrm{nm}$, $K^\text{anis} = 4000 \,\textrm{Jm}^{-3}$) where the computation time is significantly higher.

For the EQANIS model, the computation times are  almost independent of the diameter and anisotropy. This was to be expected from the series expressions in Lemma \ref{lem:seriesExansion}, since the number of evaluations scales linearly with the number of series terms $L$ 
and for a fixed $L$ thus a constant computational complexity could be considered. However, we can observe a slight increase for larger $D$, 
which is likely to be caused by non-constant computation time of the Bessel functions for different input parameters. When comparing both models with respect to the computation time, one can see that the rounded speedup for the EQANIS model varies between 65 and 3810 with a mean value of 457.

\subsection{Experiments}

\begin{figure*}
    \centering
    \includegraphics[width=1.0\textwidth]{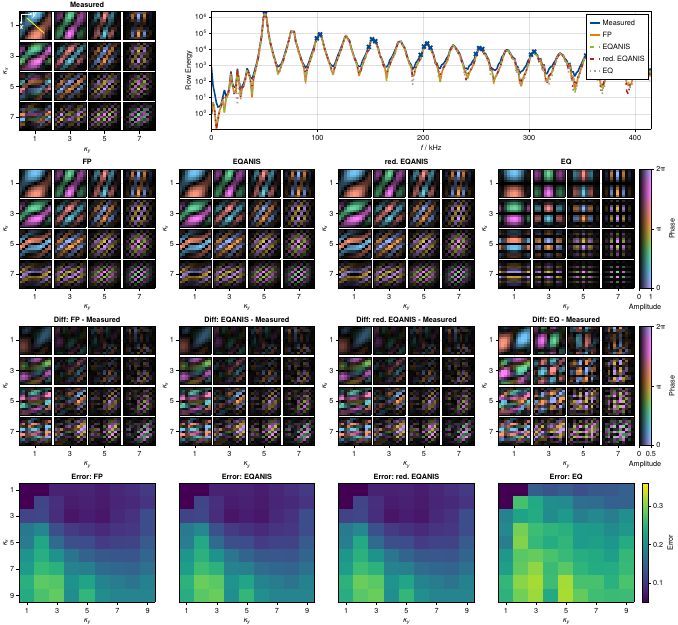}
    \caption{Comparison of the modeled and measured system matrices for the immobilized and axis-aligned ($45^\circ$) case. In the first and second row, selected frequency components of all measured and simulated system matrices are shown for mixing factors $\kappa_x, \kappa_y \in \{1,3,5,7\}$ and the $y$-receive channel using a complex colormap. Each frequency component is normalized to its maximum value. The easy axis is shown as a yellow line in the frequency component $\kappa_x = \kappa_y = 1$ of the measured system matrix.  The upper right graph shows the row energy (i.e. the $\ell_2$ norm of each frequency components) for all system matrices. In the third row, difference maps between the measured and the modeled frequency components are shown. The colormap is scaled by a factor of 2 in this case to amplify the differences. In the last row, the error $\varepsilon^\text{SM}$ is shown for $\kappa_x, \kappa_y \in \{1, \dots, 9\}$. }
    \label{fig:SMVisuImmobilizedOrientated}
\end{figure*}

For the immobilized and oriented measurement (first data set), suitable particle parameters of the models with anisotropy were found to be $D = 19\,\text{nm}$ and $K^\text{anis} = 1400\,\textrm{Jm}^{-3}$. For the EQ model, we kept the same particle diameter. In the upper part of Fig.~\ref{fig:SMVisuImmobilizedOrientated} selected frequency components (rows of the system matrix) at mixing factors $\kappa_x, \kappa_y$ of the $x$-receive channel are shown. We visualize the complex data using a color-coded 2D colormap introduced in \cite{albers2022modeling}. 
In the first row the measured system matrix is shown and in the second row, the modeled system matrices.
In general, we see that the frequency patterns are consisting of oscillating functions, where the oscillation degree depends on the mixing factors. For the immobilized and oriented case, the wave hills are merging in the direction being orthogonal to the easy axis. This effect has been intensively studied in \cite{albers2022modeling} and can be used to determine the orientation \cite{moddel2021estimating} when performing a multi-contrast reconstruction. Comparing the system matrices, we see that the FP matrix is very similar to the measured one. Differences can be only seen in some of the frequency components, highlighted in the difference plots in the third row of Fig.~\ref{fig:SMVisuImmobilizedOrientated}. Moreover, when comparing the FP model to the EQANIS model and the reduced EQANIS, there is no visually perceptible difference at all, not even in the dedicated difference plots. This was to be expected from the simulations performed in the previous section and  considering the rather small particle parameters. Taking a look at the system matrix calculated with the EQ model, we clearly see that the features caused by the easy axis aligning cannot be modeled. These qualitative findings are quantified by calculating the error $\varepsilon^\text{SM}$, which is shown for all models in the lower part of Fig.~\ref{fig:SMVisuImmobilizedOrientated}. The error has a mean of 0.136 for FP, EQANIS, and the reduced EQANIS model. For the EQ model, the error is higher and has a mean of 0.220. 

\begin{figure*}
    \centering
    \includegraphics[width=1.0\textwidth]{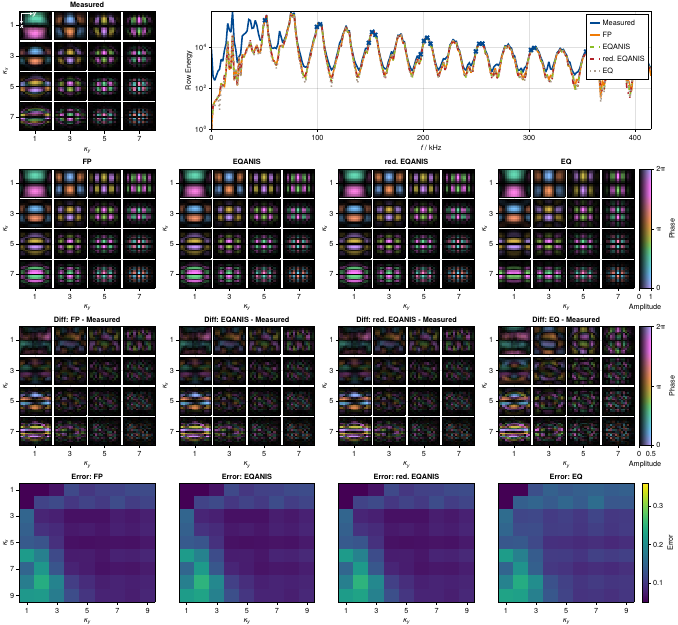}
    \caption{Comparison of the modeled and measured system matrices for the fluid case. In the first and second row, selected frequency components of all measured and simulated system matrices are shown for mixing factors $\kappa_x, \kappa_y \in \{1,3,5,7\}$ and the $y$-receive channel using a complex colormap. Each frequency component is normalized to its maximum value. The upper right graph shows the row energy (i.e. the $\ell_2$ norm of each frequency components) for all system matrices. In the third row, difference maps between the measured and the modeled frequency components are shown. The colormap is scaled by a factor of 2 in this case to amplify the differences. In the last row, the error $\varepsilon^\text{SM}$ is shown for $\kappa_x, \kappa_y \in \{1, \dots, 9\}$.}
    \label{fig:SMVisuFluid}
\end{figure*}

For the fluid measurements (second data set) we found matching particle parameters at $D = 19\,\text{nm}$, $\Kanismax = 3500\,\textrm{Jm}^{-3}$, and $q=2$. The resulting system matrices are compared in Fig.~\ref{fig:SMVisuFluid}. Again, we used the same diameter for the EQ model. Since the easy axes are now directed radially around the center, the merging of the wave hills occurs along circles. The effect is stronger in outer regions because the anisotropy increases with the distance to the center. The qualitative findings for modeling the fluid data are the same as for the immobilized and orientated data. Again, the three models accounting for anisotropy look identical and closely resemble the measured system matrix. The EQ model shows larger differences to the measurements. This is also observable in the error, which has a mean of 0.100 for the first three models and a mean of 0.121 for the EQ model.

\begin{figure}
    \centering
    \includegraphics[width=0.49\textwidth]{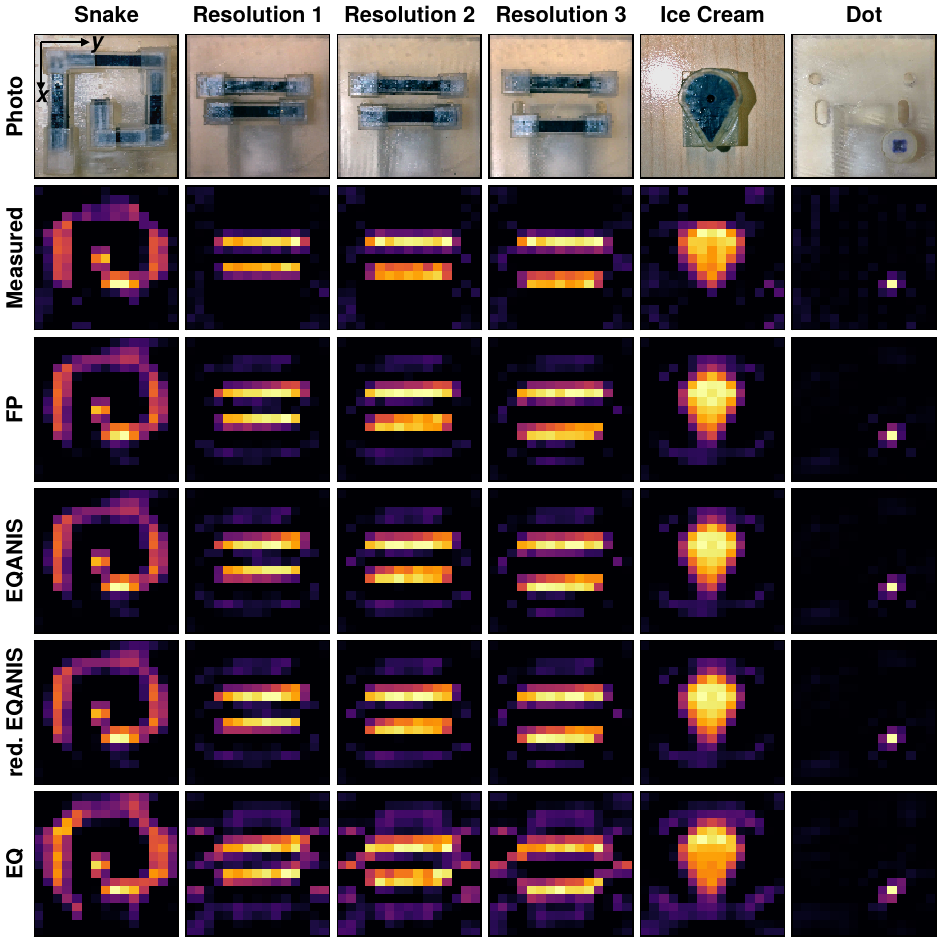}
    \caption{Reconstruction results for six different phantoms shown in the individual columns. The first row shows a photo of the filled phantom. Rows 2 -- 6 show the reconstruction results when using the measured system matrix and the four differently modeled system matrices (FP, EQANIS, reduced EQANIS, EQ).}
    \label{fig:RecoFluid}
\end{figure}

Finally, we consider the reconstruction results of the six different phantoms,  shown in Fig.~\ref{fig:RecoFluid}. The main features of the different phantom can be indicated using any of the system matrices. Quality-wise the FP-based reconstruction looks very similar to the reconstruction from the measured system matrix. Only the artifact level in the background is slightly larger and the rods of the resolution phantom look slightly less blurred. When comparing the FP reconstruction with the EQANIS and the reduced EQANIS reconstruction, there is almost no visually observable difference. This was to be expected as the difference in the simulated system matrices is so small. When taking a look at the result obtained by the equilibrium model, we see a clear decrease in image quality. First of all the background artifacts are much stronger. Moreover, there are stronger spatial deformations of the reconstructed particle concentration in the snake phantom. This can be explained by the deformed patterns in the system matrix itself.

\section{Discussion}

The results show that the EQANIS model is very promising for modeling the particle physics in MPI and enables accurate model-based reconstruction with low computational effort for model evaluation. For non-relaxing particles with  a sufficiently small diameter/anisotropy, the EQANIS model has a high similarity to the FP model. From Fig.~\ref{fig:Accuracy1D} one can derive that the anisotropy needs to be below $1500\,\textrm{Jm}^{-3}$ for $D=24\,\textrm{nm}$ diameter, below $3500\,\textrm{Jm}^{-3}$ for $D=20\,\textrm{nm}$, and below $10,000\,\textrm{Jm}^{-3}$ for $D=16\,\textrm{nm}$ diameter in order to reach an error of $1\%$. It should be noted that the FP model acts as a reference model, although it itself only has a certain accuracy since it is numerically derived from a stiff differential equation, which is challenging to solve.
In addition, Fig.~\ref{fig:Accuracy1D} will be helpful for other MPI researchers to evaluate whether the EQANIS model is suitable for their purposes.

An important advantage of the EQANIS model over the FP model is that it is an explicit model and therefore does not require the solution of a differential equation to evaluate it. As a result, the evaluation of the EQANIS model is much faster. In our experiments and with the fast series-based implementation derived in this work, we observed a speedup of 2-3 orders of magnitude. The speedup would be even higher if the accuracy for solving the FP equation is increased, which depends non-linearly on the accuracy. In addition, we observed that the solver also depends on the shape of the applied fields as well as the concrete particle core diameter, anisotropy constant, and easy axis angle. Thus, it is difficult to predict how long the solution will actually take. In contrast, the EQANIS model requires an almost constant computation time and changes only slightly because the Bessel functions require different times for different input parameters.

When looking at the experimental 2D Lissajous data, we  observed that the EQANIS model can basically replace the FP model, since the found effective particle parameters were sufficiently small implying a sufficiently small approximation error as discussed before. Note that this is likely to be caused by a specific characteristic of 2D/3D Lissajous trajectories, which constantly change the excitation angle and thus do almost never excite along the easy axis direction (colinear easy axis). In contrast, in case of a 1D excitation, the fluid particles align their easy axis along the excitation direction and in turn relaxation effects become visible for most particle systems \cite{Graeser2015}. In those cases the EQANIS model alone would not be suitable. However, one could combine it with the Debye model, which is commonly used to model 1D MPI systems. Note that in the present work we fit the transfer function and thus already implicitly include the Debye model as it is basically modeled by a convolution in time.
For 2D/3D Lissajous trajectories, the transfer function can only model global relaxation, whereas one expects  only local relaxation happening when the FFP crosses the center of the FOV.

The model-based reconstruction results are much closer to the calibration-based reconstruction results. This is remarkable since they still were slightly worse in \cite{Kluth2019}. The reasons for this can be manifold. First of all, we used a  more fine-grained parameter search since our computing hardware was much faster and also the FP implementation got more efficient. But one other important aspect is that we use a slightly improved reconstruction scheme, by applying diagonal whitening before reconstruction. This strongly increased the robustness of reconstruction and allowed us to completely mitigate a frequency selection.
This observation also raises the question of a clear distinction between the influence of the background signal and the particle model on image reconstruction.

One still open problem is the field of parameter identification. In this work we only use monodisperse particle models with a single diameter and anisotropy. Improvements can be expected by switching to the polydisperse setting as we have done for non-aligned immobilized particles in \cite{Albers2023IWMPI}. Here, the EQANIS model can play an important role because a much larger parameter space can be covered in the same amount of time. Furthermore, it is possible to calculate analytical derivatives explicitly with respect to desired input parameters, which provides the opportunity to further improve the optimization process. 

In addition to the EQANIS model, we also introduced a reduced form, which allows us to express the entire imaging operator as a product of a first operator containing certain tensor products of Chebyshev polynomials and a second operator containing a spatially varying convolution. The model reduction is to limit the number of Chebyshev tensor product basis functions to be included in each frequency components to one in \eqref{eq:sysFuncCompAnisoReduced}, which in turn allows for direct inversion of the operator as outlined in \cite{droigk2022direct}. We have found that the approximation being made is in practice negligible, which has been previously discussed in \cite{maass2020novel} and \cite{droigk2022direct} for the EQ model. Exploiting the reduced EQANIS model allows for direct reconstruction, which was initially derived and shown in \cite{droigk2023adaption}. We note that this would not be possible for the FP model, which shows that the EQANIS model is not only faster to evaluate but also has the advantage that it can be better utilized during reconstruction. Compared to the EQ model, which shares these properties, the EQANIS model is much more accurate.

More recently, the Fourier neural operator (FNO) \cite{Knopp2023IWMPI} has been proposed as a neural network approach to solving the FP equation. The FNO significantly accelerates the numerical solution of the corresponding FP equation after training the network. However, it still requires the generation of training examples, which have to be generated using classical numerical solvers, with all the associated limitations. Furthermore, the numerical quality of the FNO has not yet been sufficiently validated and its accuracy is therefore unclear. In addition, unlike the EQANIS model, the FNO model does not provide such an easily decomposable imaging operator, so that no direct image reconstruction can be derived. However, the EQANIS model is limited to non-relaxing particles, which the FNO model can include. Thus, depending on the particle and sequence type, the FNO and EQANIS models may be more appropriate.

\section{Conclusion}

We have found that the EQANIS model has great potential for model-based reconstruction in MPI using multidimensional Lissajous sequences. It can be computed efficiently and yet is accurate enough to describe typical features observed in measured MPI system matrices. It could play an important role in eliminating the need for the \mbox{$\Delta$-sample} based calibration procedure. The main limitation of the model is that one needs to be aware that it cannot describe relaxation effects, which can become significantly relevant for particular excitation sequences such as 1D excitations.

\IEEEtriggeratref{46}
\bibliographystyle{IEEEtran}
\bibliography{arXiv}

\clearpage

\section*{Supplemental Material A}

In this supplemental material, Lemma \ref{lem:seriesExansion} is proven. For simplicity, let $a=\beta|\tilde{\vek{H}}|_{12} $, $b = \beta \tilde{H}_3 $, and $c = \ak$.
The Cauchy product between the Maclaurin series of $\eUnit^{c\xi^2}$ and $\cosh(b\xi)$  is 
\begin{align}
\cosh(b\xi)  \eUnit^{c\xi^2}&= \sum_{\ell=0}^\infty \sum_{k=0}^\ell \frac{(b\xi)^{2k}}{(2k)!}\frac{(c\xi^{2})^{\ell-k}}{(\ell-k)!} \label{eq:SeriesCoshComplexGauss}\\
&=\sum_{\ell=0}^\infty \xi^{2\ell} \underbrace{\sum_{k=0}^\ell  \frac{b^{2k}}{(2k)!}\frac{c^{\ell-k}}{(\ell-k)!}}_{=d_\ell}. \nonumber 
\end{align}

If $c\neq 0$, the right hand side of the second equal sign $d_\ell$ can be further expressed by
\begin{equation}
	\begin{aligned}
		d_\ell	&= \sum_{k=0}^\ell  \frac{b^{2k}}{(2k)!}\frac{c^{\ell-k}}{(\ell-k)!}= c^\ell \sum_{k=0}^\ell \frac{b^{2k} c^{-k}}{(2k)!(\ell-k)!} \\
						&= c^\ell \sum_{k=0}^\ell \frac{\left(\frac{b^2}{c}\right)^k}{(2k)!(\ell-k)!}= c^\ell \sum_{k=0}^\ell 4^k4^{-k}\frac{\left(\frac{b^2}{c}\right)^k}{(2k)!(\ell-k)!}  \\
						&= c^\ell {\sum_{k=0}^\ell 4^k\frac{\left(\frac{b^2}{4c}\right)^k}{(2k)!(\ell-k)!}} 
      =\frac{c^\ell\sqrt{\pi}}{\Gamma(\tfrac{1}{2}+\ell)}L_\ell^{(-\frac{1}{2})}\mathopen{}\left(-\frac{b^2}{4c}\mathclose{}\right), \label{eq:SeriesTermCoshComplexGauss}
	\end{aligned}
\end{equation}
where  $L^{(\alpha)}_n(\xi)$ is the generalized Laguerre polynomial with degree $n\in\mathbb N_0$ and order $\alpha >-1$. In case $c=0$, one has $d_\ell	= \tfrac{b^{2\ell}}{(2\ell)!}$.

The Cauchy product of the Maclaurin series of the functions  $\eUnit^{c\xi^2}$,  $\sinh(b\xi)$, and $\xi$ is
\begin{equation}
	\begin{aligned}
		\xi\sinh(b\xi)  \eUnit^{c\xi^2}&= \xi \sum_{\ell=0}^\infty \sum_{k=0}^\ell \frac{(b\xi)^{2k+1}}{(2k+1)!}\frac{(c\xi^{2})^{\ell-k}}{(\ell-k)!} \\
		&=\sum_{\ell=0}^\infty \xi^{2\ell+2}\underbrace{\sum_{k=0}^\ell  \frac{b^{2k+1}}{(2k+1)!}\frac{c^{\ell-k}}{(\ell-k)!}}_{=e_\ell} 
  \label{eq:SeriesSinhComplexGauss}
	\end{aligned}
\end{equation}
further, $e_\ell$ can be simplified by
	\begin{align}
		e_\ell &= \sum_{k=0}^\ell  \frac{b^{2k+1}}{(2k+1)!}\frac{c^{\ell-k}}{(\ell-k)!} = bc^\ell  \sum_{k=0}^\ell  \frac{b^{2k} c^{-k}}{(2k+1)!(\ell-k)!}\nonumber \\
					 &= b c^\ell \sum_{k=0}^\ell  \frac{ 4^k}{(2k+1)!(\ell-k)!}\left(\frac{b^2}{4c}\right)^k \nonumber \\
      &=\frac{bc^\ell \sqrt{\pi}}{2\Gamma(\tfrac{3}{2}+\ell)}L_{\ell}^{(\frac{1}{2})}\mathopen{}\left(-\frac{b^2}{4c}\mathclose{}\right)
					 \label{eq:SeriesTermSinhComplexGauss}
	\end{align}
	under the assumption of $c\neq 0$. 
In the case of $c=0$ one has
		$e_\ell =  \tfrac{b^{2\ell+1}}{(2\ell+1)!}.$
		
Now the integrals involved in \eqref{eq:IntExpressZ}, \eqref{eq:IntExpressZ3}, and \eqref{eq:IntExpressZ12} can be integrated stepwise by interchanging the summation of the series and the integration, which is allowed since all the functions involved are analytic functions and the series terms are absolutely convergent in the entire interval $[0,1]$ to be integrated.

For the integration we use the integral formula from \cite[§11.4.10]{Abramowitz1964} for (modified) Bessel functions of the first kind
\begin{equation}
	\begin{aligned}
		\int_{0}^1 I_0&\mathopen{}\left(a\sqrt{1-x^2}\mathclose{}\right)x^n \dt[x] \\ &\overset{\dagger}{=}\int_0^{\frac{\pi}{2}} I_0\mathopen{}\left(a\sin(\alpha)\mathclose{}\right)\cos^n(\alpha)\sin(\alpha)\dt[\alpha] \\
		&= 2^{\frac{n-1}{2}} \Gamma\mathopen{}\left(\frac{n+1}{2}\mathclose{}\right) a^{-\frac{n+1}{2}} I_{\frac{n+1}{2}}(a)
		\end{aligned}
		\label{eq:IntFormulaBesselFunc}
\end{equation}
with $n \in\mathbb N$. The equality denoted  by the $\dagger$ can be shown by substitution with $x=\cos(\alpha)$ and $\dt[x]=-\sin(\alpha)\dt[\alpha]$.

The essential component of the integral in \eqref{eq:IntExpressZ} is
\begin{equation}
	\begin{aligned}
			\int_{0}^1 &I_0\mathopen{}\left(a\sqrt{ 1- x^2} \mathclose{}\right) \cosh(b x) \eUnit^{c x^2  } \dt[x] \\
   & \overset{\eqref{eq:SeriesCoshComplexGauss}}{=} \int_{0}^1 I_0\mathopen{}\left(a\sqrt{ 1- x^2} \mathclose{}\right) \sum_{\ell=0}^\infty d_\ell x^{2\ell} \dt[x]\\
			& \overset{\dagger}{=}\sum_{\ell=0}^\infty d_\ell \int_{0}^1 I_0\mathopen{}\left(a\sqrt{ 1- x^2} \mathclose{}\right)  x^{2\ell} \dt[x] \\
			&\overset{\eqref{eq:IntFormulaBesselFunc}}{=}\sum_{\ell=0}^\infty d_\ell 2^{\frac{2\ell-1}{2}} \Gamma\mathopen{}\left(\frac{2\ell+1}{2}\mathclose{}\right) a^{-\frac{2\ell+1}{2}} I_{\frac{2\ell+1}{2}}(a)\\
			& \overset{\eqref{eq:SeriesTermCoshComplexGauss}}{=} \sqrt{\pi}\sum_{\ell=0}^\infty c^\ell L_\ell^{(-\frac{1}{2})}\mathopen{}\left(-\frac{b^2}{4c}\mathclose{}\right) 2^{\ell - \frac{1}{2}} a^{-\ell -\frac{1}{2}} I_{\ell+\frac{1}{2}}(a).
	\end{aligned}
	\label{eq:ProofResultZ}
\end{equation}

The essential component of the integral in \eqref{eq:IntExpressZ3} is
\begin{equation}
	\begin{aligned}
			\int_{0}^1 & x I_0  \mathopen{}\left(a\sqrt{ 1- x^2} \mathclose{}\right) \sinh(b x) \eUnit^{c x^2  } \dt[x] \\ & \overset{\eqref{eq:SeriesSinhComplexGauss}}{=} \int_{0}^1  I_0\mathopen{}\left(a\sqrt{ 1- x^2} \mathclose{}\right) \sum_{\ell=0}^\infty e_\ell x^{2\ell+2} \dt[x]\\
			& \overset{\dagger}{=}\sum_{\ell=0}^\infty e_\ell \int_{0}^1 I_0\mathopen{}\left(a\sqrt{ 1- x^2} \mathclose{}\right)  x^{2\ell+2} \dt[x] \\
			&\overset{\eqref{eq:IntFormulaBesselFunc}}{=}\sum_{\ell=0}^\infty e_\ell 2^{\frac{2\ell+2-1}{2}} \Gamma\mathopen{}\left(\frac{2\ell+2+1}{2}\mathclose{}\right) a^{-\frac{2\ell+2+1}{2}} I_{\frac{2\ell+2+1}{2}}(a)\\
			& \overset{\eqref{eq:SeriesTermSinhComplexGauss}}{=} b \sqrt{\pi} \sum_{\ell=0}^\infty {c^\ell }L_{\ell}^{(\frac{1}{2})}\mathopen{}\left(-\frac{b^2}{4c}\mathclose{}\right)2^{\ell-\frac{1}{2}} a^{-\ell-\frac{3}{2}} I_{\ell+ \frac{3}{2}}(a).
	\end{aligned}
	\label{eq:ProofResultZ3}
\end{equation}

For the essential component of the integral in \eqref{eq:IntExpressZ12} we will use that the following holds
\begin{equation*}
	\begin{aligned}
		\int_{0}^1 & \sqrt{ 1- x^2} I_1\mathopen{}\left(a\sqrt{ 1- x^2} \mathclose{}\right) \cosh(b x) \eUnit^{c x^2  } \dt[x] \\
  &= \int_{0}^1 \pnpt[a]{I_0\mathopen{}\left(a\sqrt{ 1- x^2} \mathclose{}\right)} \cosh(b x) \eUnit^{c x^2  } \dt[x] \\
		 &=\pnptn[a]{\int_{0}^1 I_0\mathopen{}\left(a\sqrt{ 1- x^2} \mathclose{}\right) \cosh(b x) \eUnit^{c x^2  } \dt[x]}.
	\end{aligned}
\end{equation*}

Consequently, we have
\begin{equation*}
	\begin{aligned}
		\int_{0}^1 &\sqrt{ 1- x^2} I_1  \mathopen{}\left(a\sqrt{ 1- x^2} \mathclose{}\right) \cosh(b x) \eUnit^{c x^2  } \dt[x] \\ &= \pnptn[a]{\int_{0}^1 I_0\mathopen{}\left(a\sqrt{ 1- x^2} \mathclose{}\right) \cosh(b x) \eUnit^{c x^2  } \dt[x] }\\
			&\overset{\eqref{eq:ProofResultZ3}}{=} \pnptn[a]{\sqrt{\pi}\sum_{\ell=0}^\infty c^\ell L_\ell^{(-\frac{1}{2})}\mathopen{}\left(-\frac{b^2}{4c}\mathclose{}\right) 2^{\ell - \frac{1}{2}} a^{-\ell -\frac{1}{2}} I_{\ell+\frac{1}{2}}(a)}\\
			&\overset{\ddagger}{=} \sqrt{\pi}\sum_{\ell=0}^\infty c^\ell L_\ell^{(-\frac{1}{2})}\mathopen{}\left(-\frac{b^2}{4c}\mathclose{}\right) 2^{\ell - \frac{1}{2}} \pnptn[a]{a^{-\ell -\frac{1}{2}} I_{\ell+\frac{1}{2}}(a)},
	\end{aligned}
\end{equation*}
where the step denoted by $\ddagger$ is allowed since all functions involved are analytic and the series is absolutely convergent. Since for the modified Bessel function of the first kind, the following differentiation relation \cite[§9.6.28]{Abramowitz1964} applies
\begin{equation}
	\dndt[\xi]{[\xi^{-\alpha} I_\alpha(\xi)]} = \xi^{-\alpha} I_{\alpha+1}(\xi) \quad\text{for } \alpha\geq 0.
	\label{eq:DevBessel}
\end{equation}
One obtains
\begin{equation}
	\begin{aligned}
 		\int_{0}^1 &\sqrt{ 1- x^2} I_1  \mathopen{}\left(a\sqrt{ 1- x^2} \mathclose{}\right) \cosh(b x) \eUnit^{c x^2  } \dt[x]\\
   &\overset{\eqref{eq:DevBessel}}{=} \sqrt{\pi}\sum_{\ell=0}^\infty c^\ell L_\ell^{(-\frac{1}{2})}\mathopen{}\left(-\frac{b^2}{4c}\mathclose{}\right) 2^{\ell - \frac{1}{2}}a^{-\ell-\frac{1}{2}} I_{\ell+\frac{3}{2}}(a).
	\end{aligned}
	\label{eq:ProofResultZ12}
\end{equation}

\noindent By combining the results \eqref{eq:ProofResultZ}, \eqref{eq:ProofResultZ3}, \eqref{eq:ProofResultZ12} with the integral representations \eqref{eq:IntExpressZ}, \eqref{eq:IntExpressZ3}, \eqref{eq:IntExpressZ12}  one obtains the series representations \eqref{eq:seriesZ}, \eqref{eq:seriesz3}, \eqref{eq:seriesz12}, which completes the proof of  Lemma \ref{lem:seriesExansion}.

\end{document}